# Competence Assessment as an Expert System for Human Resource Management: A Mathematical Approach

Mahdi Bohlouli[a,*], Nikolaos Mittas[b], George Kakarontzas[b,c], Theodosios Theodosiou[b], Lefteris Angelis[b], Madjid Fathi[a]

[a]*Institute of Knowledge Based Systems, University of Siegen, Hoelderlinstr. 3, 57076 Siegen, Germany*
[b]*Department of Informatics, Aristotle University of Thessaloniki, 54124 Thessaloniki, Greece*
[c]*Department of Computer Science and Engineering, T.E.I. of Thessaly, 41110 Larissa, Greece*

**Abstract**

Efficient human resource management needs accurate assessment and representation of available competences as well as effective mapping of required competences for specific jobs and positions. In this regard, appropriate definition and identification of competence gaps express differences between acquired and required competences. Using a detailed quantification scheme together with a mathematical approach is a way to support accurate competence analytics, which can be applied in a wide variety of sectors and fields. This article describes the combined use of software technologies and mathematical and statistical methods for assessing and analyzing competences in human resource information systems. Based on a standard competence model, which is called a Professional, Innovative and Social competence tree, the proposed framework offers flexible tools to experts in real enterprise environments, either for evaluation of employees towards an optimal job assignment and vocational training or for recruitment processes. The system has been tested with real human resource data sets in the frame of the European project called ComProFITS.

## 1. Introduction

An abundance of emerging technologies, especially in the IT sector, imposes needs for further expertise in companies. Furthermore, existing human resources might not be able to fulfill competence gaps in enterprises. To deal with this problem, companies should provide on-the-job training to their employees or recruit new experts. In any case the competence gaps should be first identified. For identifying any person's competence gaps, her/his background knowledge in connection to the target job should be properly assessed. Even in the case of the on-the-job-training, it should be properly understood, which trainings employees need in order to improve competence gaps in the company.

Hence, an efficient management of human and non-human resource competences prevents imposition of exorbitant costs, improves the quality of products and services and facilitates better workforce planning. Meanwhile the use of expert systems in Competence Management (CM) processes supports industries in making the strategic decisions, identifying unused competences, predicting future expected competences and providing better career development opportunities and improved work performance for employees (N Mittas et al., 2015).

---

\* Corresponding author. Tel.: +49-271-740-2626; fax: +49-271-740-3038.
E-mail addresses: mahdi.bohlouli@uni-siegen.de (Mahdi Bohlouli), nmittas@csd.auth.gr (Nikolaos Mittas), gkakaron@teilar.gr (George Kakarontzas), theodosiou@statnson.com (Theodosios Theodosiou), lef@csd.auth.gr (Lefteris Angelis), Fathi, fathi@informatik.uni-siegen.de (Madjid Fathi).

Usually companies have a problem in assigning the right person with the right expertise (performance) to the right position at the right time (period). The authors define this as 4Ps rule (Person, Performance, Position, Period) which is a key for an efficient person-job-fit process. Any of those Ps should be addressed, focused and assessed properly in companies. As a first step, there should be sufficient candidates (person) to fulfill the competence needs of a position. This can be done from already employed people or new candidates who seek a job in the company. Required job qualifications (performance) of those Human Resources (HRs) should be assessed through methods like questionnaires or 360-degree assessment (Hazucha, Hezlett, & Schneider, 1993). In order to identify the most performant and competitive employee/candidate, companies need to know which competences and qualifications (performance) do they need for the target job position. The difference between competences of any person and required competences for a target job position is referred to as competence gap in this work. The competence gap should be discovered as early as possible before incurring additional unplanned costs, due to the misuse, skill mismatch or lack of human resource competences.

There has been significant research on competences and Competence Management (CM) (Amiri, Zandieh, Soltani, & Vahdani, 2009; Mahdi Bohlouli et al., 2013; Mahdi Bohlouli, Ansari, & Fathi, 2012; Małachowski, Różewski, & Zaikin, 2011; Peters & Zelewski, 2007; Różewski, Kusztina, Tadeusiewicz, & Zaikin, 2011; Shavelson, 2013). Competences are often defined as knowledge, expertise, skills and abilities that people need to carry out job roles. European Commission (EC) has defined competence (European Commission, 2008) as "proven ability to use knowledge, skills and personal, social and/or methodological abilities, in work or study situations and in professional and personal development". The term competence has been defined and focused from different perspectives such as general competences (Gary Dessler, 2015), soft skills (Bailey, 2014; Robles, 2012), business skills (Bailey, 2014) and technical competences (Bailey, 2014). However, there is still a lack of clarity in the classification of differences between competence and competency. Some works in literature and reports intermix the definition of both. According to the Cambridge dictionary[†], competence is "the ability to do something well" and competency is "an important skill that is needed to do a job". Consequently, the term competence reflects the performance perspective of required skills (e.g. competency) to do a specific job. Teodoresco (Teodorescu, 2006) has referred to Gilbert's definition of the competence (Gilbert, 1978) and highlighted the worthy performance aspect as the key difference between "competence" and "competency".

Human Resource Information systems (HRIS) need a great involvement of CM components. Lindgren et al. addressed the challenge of missing studies in the Competence Management Technologies (CMTs) in HRISs (Lindgren, Henfridsson & Schultze, 2004). They defined CM as specific information systems that help organizations to manage competences in the organizational and individual levels. Similarly, Baladi defined CM as specification of the competence requirements, identification of the competence gaps as well as competence sourcing, development and staffing (Peter, 1999). Draganidis and Mentzas listed competence identification, assessment, acquisition, and usage as CM processes (Draganidis & Gregoris, 2006). Positive reactions and effects of using a Competence Management System (CMS) in organizations have been described in (Lindgren et al., 2004) from employees, managers and organization's perspectives. According to their description, employees focus on their competence development and marketing their know-how and need managers to support them. Managers are flexible, fast and more accurate in their job since they clearly understand who knows what. Finally, organizations are able to support a systematic competence development and strategic competence supply.

Many companies and organizations face serious difficulties in understanding their acquired and required competences. This results in inefficient use of human resources, since they lack the integration of scientific and computerized algorithms in the HRM processes. Mishra and Akman state that HRM still lacks in the application of the Information Technology (IT) (Mishra & Akman, 2010). According to their survey involving 206 IT managers and professional from various sectors in Turkey, IT has significant and positive improvements in all sectors in terms of HRM. Snell et al. estimate that HR-related issues of each employee cost approximately $1,500 annually for typical organizations which can be doubled and even tripled in less efficient organizations (Snell, Stueber, & Lepak, 2001). Mishra and Akman concluded from the literature review that "One of the impacts of IT is that it enables the creation of an IT-based workplace, which leads to what should be a manager's top priority-namely, and strategic competence management" (Mishra & Akman, 2010).

---

[†] Cambridge Dictionaries Online, accessed via the web address: http://dictionary.cambridge.org/, on July 2015.

In addition to discussed requirements and needs associated with CMSs, Jackson defines an expert system as a computer system imitating human-expert decision-making (Jackson, 1998). Consequently, an expert system supports decision making in complex systems, predictions, assessments or monitorings. Hyes et al. classifies expert systems in 10 categories ranging from "planning" and "design" to "instruction" and "control" (Hayes-Roth, Waterman, & Lenat, 1983). According to their description, an "instruction" type expert systems address problems such as assessment of behavioral and workforce knowledge of experts such as employees' competences as well as supporting decision making processes in prioritization of experts or recruitment systems. Similarly, expert systems for improving problem solving ability of students (Hwang, Chen, Tsai, & Tsai, 2011), capability assessments in skill-based environments (Otero & Otero, 2012), competence evaluation of firms (Amiri et al., 2009) as well as modeling and assessment of student behavior and competences in intelligent tutoring systems (Jeremić, Jovanović, & Gašević, 2012) have been proposed in this regard. Proposed mathematical competence assessment approach in this paper aims to support the requirements of an expert system in recruitment processes in the HRM departments.

The main focus of this research is first defining a reference competence tree which covers modeling of required competences in a wide variety of sectors and applications. Second, this research assessed and analyzed the qualifications of individuals based on required competences. The profiling and visual representation mechanism of this work is important from the managerial and strategic viewpoint which provides visual semantics for high level decision makers about the competence gaps in organizations. Meanwhile, a mathematical representation of competences (e.g. competence profiles), prioritization of knowledge workers with respect to the required competences, clustering them based on their qualifications as well as the person-job-fit processes are the main highlights of this research from a scientific perspective. The methods that have been developed in this research are not dedicated to the HRM area, but can be used in assessing different services or products based on the specific preferences or settings in order to choose the best solution. In general, this is called the 'required-best-fit' concept from the authors' perspective which reflects the idea of choosing the best fitting object (e.g. person or product) based on identified gaps (e.g. requirements or configuration).

For the purpose of competence assessment in organizations, two different assessments have been considered. The first is to assess people from others' (e.g. colleagues and immediate manager) perspective. This is being handled through 360-degree feedback method (Hazucha et al., 1993). The next is to assess the personal knowledge and qualifications of human resources. This is known as self-assessment in this research and has been handled through questionnaires. The final results of both assessments are being summarized as a person's profile (called acquired competence data (ACD) in this research). ACDs are being matched and compared with the competence configuration of the company or organization based on the required competences for the specific job role (called required competence data (RCD) in this research). RCD reflects the required competences defined by the needs of a new job which are defined according to the reference competence pyramid. As a result, for each person, his/her qualification for specific jobs can be analyzed and recommendations or possibilities for future improvement can be defined.

This article consists of 6 sections. In section 1, an adequate background and objectives of the research have been provided. In section 2, related research work and literature review are covered. Section 3 builds by providing further details about the Professional, Innovative and Social (PIS) competence tree. This PIS tree has been inspired from the European funded CoMaVet research project‡. In section 4, the mathematical background of the PIS tree is covered, adding the statistical and mathematical algorithms. Section 5 covers the implementation details and analysis of the results. Finally in section 6, a short conclusion and discussion on the future work are discussed.

## 2. Related Work

The term "Competence" has been defined first by White as a performance motivation (White, 1959). He described competence from a psychological point of view as a capacity of an organism to interact with the environment in an effective manner. Ennis provided a useful and exhaustive literature review (Ennis, 2008) about competence definitions and models and reviewed the needs to use the competence models and also the practical components of competence models. Because of the mobility of the workforce and also retirement, competence models are also being used for succession planning in addition to their current uses (Ennis, 2008). Lundberg referred to competence from executive planning and development perspective (Lundberg, 1972). He focused on "knowledge", "attitude" and "ability" as building blocks of executive competences which can be defined respectively as thinking, feeling and doing in terms of activities. McClelland addressed competence assessment and "modern competence movement" in his highly cited publication (McClelland, 1973).

Accordingly, Gilbert highlighted a behavioral engineering model as a major part of competence and addressed it using a theory of engineering human performance. He issued the ways to measure and assess human performance (Gilbert, 1978). In this context, an analysis of human behavior and its consequences provides an insight about valuable performance. A measure of worth can be described as a function of the value and cost, measured by following equation (Worth=Value/Cost) as suggested in (Gilbert, 1978). Therefore, worth becomes greater when there would be more value with fewer costs. People with such conditions of producing valuable results without using costly behavior are called competent people.

Sandberg focused on competence at work and has seen competence as a specific set of knowledge and skills required to perform a specific job (work) (Sandberg, 2000). Additionally, a multi-dimensional and holistic typology of competence has been argued in (Delamare Le Deist & Winterton, 2005). "Competence" has been addressed as a fuzzy concept in this tentative work through different practices in some countries, especially in the US, UK, Germany and France. An extension of competence analysis depth, an investigation of greater competence details in some occupations, as well as identification of the rift between rationalist and interpretative approaches are stated in (Delamare Le Deist & Winterton, 2005) as the main challenges of the field. It is also mentioned that developing any system that integrates those wide and disparate directions will benefit an efficient competence analytics for on-the-job-training. Badaracco and Martinez proposed item selection process using multi-criteria decision model in the Computerized Adaptive Tests (CAT) approach for competence training. Their approach enhances the accuracy of adaptation of CAT to student's competence level (Badaracco & Martinez, 2013).

Similarly, European Commission examined the literature from France, UK, Germany and USA and defined competence as a composite definition of cognitive, functional, personal and ethical competences (European Commission, 2008). As a result, competence is the use of theories and informal tacit knowledge acquired experimentally (e.g. cognitive competence), functional abilities required in a given area of work, learning or social activity (e.g. functional competence), know-how to manage special situations (e.g. personal competences) as well as ownership of specific personal and professional values (e.g. ethical competence). Self-direction is mentioned as a critical factor to define the competence level of individuals. Competence assessment of individuals based on self-direction means his/her talent in an integration of those stated competences in this definition to be used for specific challenges, goals, situations and job roles (European Commission, 2008). Huan et al. used multi-objectives evolutionary algorithms to optimize multi-criteria expansion of competence sets. The main innovation, according to their paper, is consideration of the multi-criteria perspective of competence sets (Huang, Tzeng, & Ong, 2006).

---

‡ Accessed via the web address: http://www.adam-europe.eu/prj/3962/project_3962_de.pdf, on January 2015.

Draganidis and Mentzas reviewed 22 commercial CMSs and 18 Learning Management Systems (LMSs) and provided further recommendations for research and development in this erea (Draganidis & Gregoris, 2006). He defined competence identification, model, assessment, management, standard and profile as essential components of any Competence Management (CM) system. His conclusion based on the examination of different systems is that competences are important in workforce planning, recruitment management, learning management, performance management, career development and succession planning. Further research on standards such as XML, W3C and RDF, semantic technologies, CM systems with self-service support are recommended (Draganidis & Gregoris, 2006). Strohmeier and Piazza reviewed intensively an increasing number of data mining projects and publications on the subject of human resource management. They also provided recommendations for further research in this area. According to their recommendations, systematic overview of functional HR application areas, evaluation and customization of existing or developing new data mining methods in the HR area as well as supporting the specification and provision of suitable HR data are areas that need further research (Strohmeier & Piazza, 2013).

Baladi (1999) described processes and components of new established knowledge and competence management initiative at Ericsson. A web-based CM application development at Ericsson supports individual and organizational CM and considers "analysis of future requirements", "analysis of the present situation", "gap analysis" and "sourcing of competences" as major CM processes (Peter, 1999). A CM system in an organizational level requires intensive interaction with Knowledge Management (KM) and in fact enterprises should be aware of 'who knows what?'. Baladi also similarly addressed the gap analysis as a function of identifying the gaps between acquired and required competences (Peter, 1999). Identification and exploring core competences in different sectors and areas have been studied in (Lee, 2010; Lin, Yang, Kang, & Yu, 2011; Wu, 2009). These works cover required professional competences for areas of integrated circuits design services, core competences of R&D technical professionals as well as high performers.

Bailey studied the importance of non-technical Knowledge, Skill and Ability (KSA) in a successful technical world with main focus on the IT sector (Bailey, 2014). The motivation for her research, as she described, was estimates of a need for one million new IT workers in 2018 and 1.4 million IT job openings by 2022 in the US. Consequently, she identified essential non-technical competences in IT sector in order to prepare responsive university curriculum in the next steps. Through a survey study of collecting necessary information from different sources, she identified 32 desirable non-technical, 12 business and 20 soft skills. The conclusion was that many computer degrees have a general curriculum in order to prepare candidates for a wide variety of IT jobs. At the same time, some IT companies hire candidates with less technical competences, but more competent in soft and business skills. As a result, the recommendations in this paper could be the base for customizing of computer science studies in order to prepare more competent candidates for the industry needs (Bailey, 2014). Shavelson used statistical models in the competence assessment. He proposed six facets for the competence, which identify the domain that the competence measurement should be developed (Shavelson, 2013).

Amiri and colleagues (Amiri et al., 2009) proposed a hybrid multi-criteria decision-making model for firms' competence evaluation. The main focus of their work is on the competence evaluation of firms rather than human resources. Their competence evaluation process consists of the following steps: (1) Measuring the weight of each criterion with Adaptive Analytic Hierarchy Process (AHP) Approach (A3). (2) Appraise the performance of firms using Fuzzy set theory. (3) Transform Fuzzy numbers into interval data using α-cut (4) Ranking the firms' interval data because of the different values for α. (5) Apply Linear Assignment to obtain final rank for alternatives (Amiri et al., 2009).

Numerous funded research and practical projects focused on innovative ideas with scientific and industrial use of the CM in their practical and strategic results. Some of these projects concentrated on the application of CM in specific fields such as agriculture, business administration, mechanical engineering and vice versa, and provide a model or system which works only in an associated sector. The main scientific part of such projects depends on a domain specific competence model associated with ontology mapping. However, they lack an easy and fast adoption to other sectors and also generalization aspect. Some others focus on theoretical aspects of the CM and provide strategic recommendations and reports rather than a practical result. These projects collect the views of Human Resource Management (HRM) experts by well-designed surveys and provide an analysis of survey results indicating the HRM challenges and future road-maps. A summary of selected research projects is given in the following. Generally, provided information on the official website of those projects is used as a source of knowledge about their description in the following.

The German funded research project, "Confidence Competence Management as a System for Balancing Flexibility and Stability Needs (CCM2)"[§], uses an integrated trust and competence management to balance the flexibility and stability of companies for change and innovation potentials. This is done by developing a web based toolbox for balanced HR policies. Similarly, the "Dynamic Interdependency of Product and Service in Production Area" project[**] focuses on the industrial product service sector. The personnel competences in heterogeneous work systems are stated as an important enabler to social actors for performing successfully. Therefore, the project is based on the specification, measurement and development of personnel competences.

The "Business Simulation Game for HLB (Hybrid service bundles) specific skills development" project[††] aims at developing a prototype for business game. The project focuses in the hybrid power sector and tries to optimize VET with providing an insight about later required competences. Facing demographic challenges that may arise from a strategic deficit is a key point for "competence-oriented corporate coaching for sustainable competence management in SMEs (4C4Learn)" project[‡‡]. This project supports SMEs in developing occupational competence models and use of it for intra- and inters- company technical and demographic challenges. The "Modeling and Measuring Competences in Higher Education (KoKoHs)" is a funding initiative[§§] that consists of 24 research projects in different fields to cover a wide competence area in higher education. All sub-projects in the frame of this very big project focus on assessment and modeling of teaching competences in different fields such as mechanical engineering, mathematics, business administration, medical science, informatics and many more.

Furthermore, the "Competence Models for Assessing Individual Learning Outcomes and Evaluating Educational Processes" project is a type of priority program[***] funded through DFG that is a collection of 30

---

projects. Participating experts with a cognitive orientation in specific disciplines cover a wide variety of ranges provided further research results on how to test and train competences in their respected field. The main focus of "Pedagogical Knowledge and the Acquisition of Professional Competences in Training for Teachers" project[†††] is on improving a training of teachers. A basic hypothesis is "Education Scientific contents and contexts represent a conceptual framework, need the teachers to interpret classroom and school events appropriately to reflect and thus to be used for managing occupational requirements."

A computer-based test method developed in "Measuring Experimental Competences in the Large Scale Assessments" project[‡‡‡] allows valid and reliable measurement of experimental competence comprehensively. "Technology-based Assessment of Skills and Competences in VET (ASCOT)"as an initiative[§§§] consists of 21 projects which form six main areas and need close cooperation between research institutions, VET practitioners and facilities. Competence modeling and detection through simulated work environments is the main hypothesis in this initiative. An "Assistance System for Demographics Sensitive Company-Specific Competence Management for Production and Logistics Systems of the Future (ABEKO)" project[****] focuses on specific fields of production and logistics.

Table 1. Comparison of indicative works from the literaturewith the proposed approach from two different perspectives: (1) flexibility and (2) integration of computer science methods

| Literature/ Approach | Flexibility and Integration of computer science methods |
|---|---|
| Badaracco & Martinez, 2013 | This work uses computer science methods, namely multi-criteria decision making, but has shortcomings in adopting to a wide variety of sectors and competences areas. As a result, this approach is not applicable to different sectors. Meanwhile, it adopts tests to students' competences, but it doesn't support competence analysis towards matching and competence gap fulfillment. |
| European Commission, 2008 | This work examines the literature towards enriching the competence definition and covering all potential criteria in competence management, but lacks practical implementation of the concept. As a result, this work provides very good theoretical background for our research area. Integration of computer science methods in this regard will add further key advantages to the outcomes of this research. |
| Huang, Tzeng, & Ong, 2006 | This work uses computer science methods and algorithms, but lacks the flexibility and adoption to a wide variety of careers. |
| Technology-based Assessment of Skills and Competences in VET (ASCOT)" | This initiative as a collection of multi funded research projects is a good sample of applying competence management in the VET and learning activities. As described in the paper, this initiative covers a lot of educational skills and also contributes to assessment phases. Most of projects in this initiative are theoretical studies lacking application of computer science methods as well as they are limited to the educational skills, lacking an application to other sectors. Meanwhile, these projects focus only on the assessment phase, rather than matching and sorting of job seekers and competence-based employee selection. |
| Amiri et al., 2009 | This approach uses computer science methods, namely AHP algorithm in multi-criteria decision support, in competence management. The approach provides practical implementation results as well. It lacks however, the generalization and flexibility to a wide variety of sectors and it lacks proposing any competence model. Also the application area differs than this work since it is applied to firms' competences rather than individuals. |

---

| Lee, 2010; Lin, Yang, Kang, & Yu, 2011; Wu, 2009 | Both works studied different perspectives of the competence management, but lack practical implementation and application of integrating computer science methods. |
|---|---|

The proposed approach in this paper supports the requirements of CMSs discovered from the reviewed common definitions of competence and CM in this section. In particular, these requirements are considered in setting up a reference competence tree proposed in Section 3. This competence tree, as a general hierarchical model, is used in competence assessment of employees/candidates in various careers in order to collect acquired competence data. All in all, an intensive literature review in this section clarifies that CMSs demand scientific and applied mathematical methods to improve job performance in organizations. As shown in the literature review, earlier efforts in this area are mainly from psychological and HRM points of view which depend on paper-based and traditional processes and lack integration of computer science methods and accordingly computerized CMSs. Furthermore, reviewed definitions of key publications about competence and CM results in recognition of the dimensions and the hierarchy of the reference competence tree. In this way, this tree supports a wide variety of competence definitions and job requirements in various careers.

Moreover, it is concluded from reviewed literature that any CMS should consist of at least: (1) identification of required competences (competence discovery), (2) assessing acquired competences (competence assessment), (3) matching acquired and required competences (competence analysis), and (4) providing further competence development plan as well as recommendations for improving the competence gaps (competence development). Most of reviewed research projects are on the basis of ontologies or traditional paper (or interview) based competence assessment methods. As stated earlier, these approaches face difficulties in an easy adaptation of such approaches to new sectors (careers). In general, 86% of studied funded research projects are in one specific career such as nursing, teaching or politics without having any insight on the extension of their approach or providing any flexibility to further careers and case studies.

Furthermore, the main challenge and limitation that similarly addressed in most of the literature is a lack of efficient and generalized competence matching method. Considering the European refugee crisis and the main problem of skill-mismatch, this work efficiently facilitates e-recruitment in a form of skill discovery and matching solution. As stated earlier, the proposed reference competence tree provides easy adoption to further sectors and job descriptions due to the mathematical background and representation of domain specific competences with numerical values and differentiating them by defining weights for various career sectors. One can also add or remove further competences categories to the proposed tree. In the case of using ontologies for domain specific competence analysis as of reviewed research projects, customization of already developed domain ontology to the new sector demands same efforts as developing the new one. Because, it strongly depends on the context of desired sector. This difficulty has been facilitated in proposed approach of this paper by proposing a mathematical representation and analysis of competences.

## 3. The Reference Competence Tree and Assessment

As stated earlier, the proposed competence reference model in this research is called Professional, Innovative and Social (PIS) competence tree which has been inspired from CoMaVet project and composed of 3 tiers (see Figure 1). Each tier has a main level-1 competence category (e.g. PIS competences) and 4 level-2 sub-categories, followed by 16 level-3 sub-sub-categories (see figure 1). This hierarchical structure facilitates an integration and configuration of different assessment methods such as 360 degrees feedback method or self-assessment questionnaires in entire or parts of the PIS tree. The questions that should be configured for any of the PIS categories are called statements in this work. Each node can have one or more statements with weights. The HRM department in organizations initiates the PIS tree and defines affiliated weights for all nodes. The statements are normally being defined only for level-3 competences. But it is conceptually possible to assess higher level competences through tests and questionnaires as well. We assume to assess level-3 competences which are the lowest level in the tree and calculate higher levels (e.g. level-1 and level-2 competences) through associated algorithms.

Each node in the PIS tree has a weight. In some literature and other standards (Antony, 2003), a pyramid representation is being used instead of the tree. The objective of using a tree in this research is to provide full flexibility to the users to set an ad-hoc competence configuration with different weights and levels of the importance. Due to the fact that the structure and strategies of different companies and sectors differ from each other and they do not use a unique and common competence model, and also required competences are not equal for different job roles, we prefer not to give an importance ourselves to any of those nodes in the PIS tree. As a result, in a raw PIS tree, all nodes have equal weights by default which should be changed by key decision makers of any company depending on their strategies and target field. The hierarchical architecture of the competence tree is provided in Figure 1.

As stated earlier, a full assessment process consists of two independent phases as (1) multi source assessment, and (2) self-assessment. Both of those methods use the reference PIS competence tree as a source of organizations' competence configuration. Depending on the pre-configured PIS tree, heads of departments provide statements for the self-assessment and 360 degree assessment methods. The statements are grouped and correspond to specific questionnaires. Questionnaires are defined ad-hoc and cover one or more specific level-3 competences. In general, each questionnaire can comprise different corresponding statements, which are selected from the statement repository or entered individually. The results collected through both assessment methods are being summarized in the employee profile as ACD (Acquired Competences Data) for each level-3 competences. The HRM department or domain experts (e.g. head of department) should prepare associated questionnaires and tests (e.g. "statements") for those competences at level-3. In the system, various tasks can be assigned to various roles. The current assignment of tasks, such as statement creation, is in relation to the requirements of existing industrial partners in this research and may be different for different companies. However, this does not represent a problematic area for the system, since in the system the definition of security is declarative and does not require significant effort to alter. Therefore, the case of the head of department is an example in this case and different roles can undertake these responsibilities in other companies with only minor modifications in the system in the future if required.

The immediate manager decides often in cooperation with the HRM department on the initiation of an assessment activity for a specific employee (an assessee), selects competences from the competence tree that should be assessed in relation to the specific job (e.g setting up the job profile description) and initiates relevant questionnaire with associated statements. Additionally, the manager sets up a detailed competence assessment plan, including time schedule, assessment team and definition of questionnaire with statements. The manager should also define the weights of different assessment types that are being handled in the assessment phase. The value (acquired competence data, ACD) of each sub-sub-category in the level-3 needs to be summarized using Equation (1), where $W_a$ indicates a weight that has been given to specific assessment type.

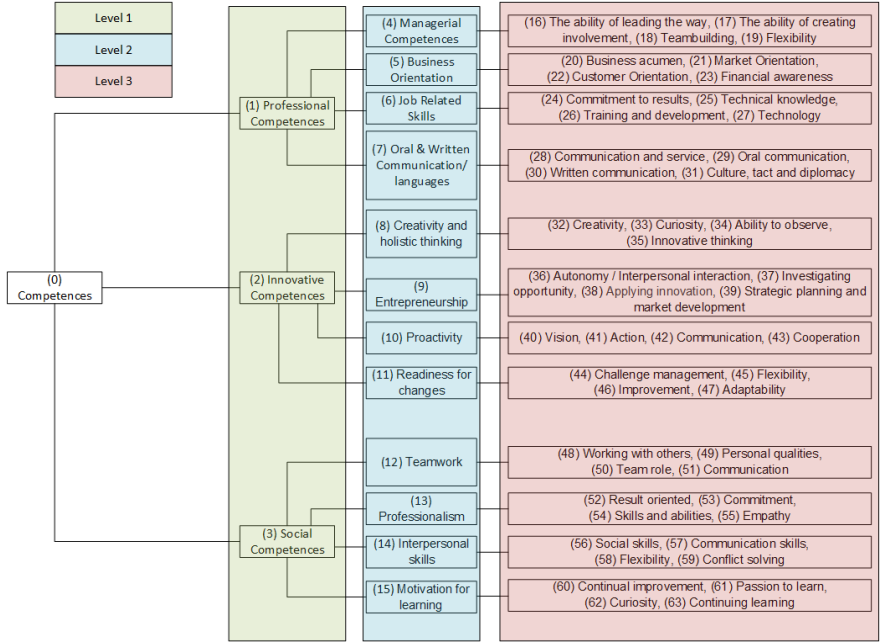

**Figure 1.** The hierarchy of PIS competence tree (the Hierarchical Cumulative Voting (HCV) in the mathematical algorithms will be based on the structure of this tree)

This is an integer number that applies for whole categories in the PIS tree of the same assessment. Similarly, the $V_{i,j,k}^{at}$ stands for values achieved in the assessment type *at* for a level-3 competence of *i,j,k*. This equation is being used only for level-3 competences and once the values of any sub-sub-category is finalized, they are being used for calculating higher levels. The total number of assessments is being defined as *n*. About any of the values in the assessments and how are the values being achieved for any of those methods is the topic of the next section.

$$ACD_{i,j,k} = \sum_{at=1}^{n} W_{at} \times V_{i,j,k}^{at} \qquad (1)$$

The level-1 categories in the competence tree are professional, innovative and social competences. This level is the highest level consisting abstraction of lower layers' measurements. For example, assessment results of 16 sub-subcategories from $C_{1.1.1}$ to $C_{1.4.4}$ are summarized (using a weighted mean calculation) as a $C_1$ in the level 1. This abstraction follows a hierarchical architecture of the tree. Interpretation of competences at this level provides an insight about general strengths and weaknesses as well as collective competence gaps of enterprises. As an example, facts such as 'An enterprise X lacks, in general, social competences.' can be achieved by analyzing the values at this level. However, a visualization of level-2 competences grasps inner competence needs of an enterprise. This is in mutual connection to enterprises' policies and requirements. Identification of social competences as a major competence gap in the aforementioned example, does not efficiently provide detailed information about enterprises' competence gaps. Emphasizing social competences as a major competence gap should evolve more detailed competence configuration as well as enterprise specific strategic and cultural needs. As a result, setting up a job description based on the level-2 sub-categories are the optimum solution that fulfills most of competence gaps.

## 3.1. Multi Source Assessment

The multi source assessment uses the 360-degree method to collect the feedback from the employee's immediate circle at a company or an organization. This method is helpful in collecting the viewpoint of colleagues who work closely with an assessee at the same level and department. The results of the 360-degree feedback assessment are stored in the system in the form of numerical values. These numerical values are qualitative facts and figures like to rate an assessee's competence level for an associated item in the PIS tree.

The 360-degrees feedback and self-assessment results for each assessee are stored in his competence matrix. A 360-degree feedback method normally involves an assessee, three colleagues and the immediate manager (the assessment team). A weighted arithmetic mean of the results collected from all participants in the assessment team for a specific assessee is stored as a person's competence matrix as raw data. The results of an assessment activity in level-3 competences are basically presented as numeric values from the Likert scale (Likert, 1932).

In the 360-degree feedback method, it is the head of department, who initiates the assessment team and manages the assessment procedure. In this assessment phase, the lowest data level presents the values from the questionnaire generated from answers of the statements for a specific competence in level-3. Normally, there are four sub-sub competences (level-3) attached to one overlying sub competence in level-2. As an example, "Teambuilding", "The ability of leading the way", "Flexibility" and "The ability to create involvement" are four sub-sub-competences in the level-3 that are all attached to the overlying sub-competence "Managerial competences" on level-2. Any of four sub-sub-competences in the level-3 is assessed through four statements. A weighted average calculation of the four sub-sub-competence values at level-3 provides the competence value of, for instance, 'managerial competence' at level-2. From the assessment activity, there are now 16 values from the Likert scale in the level-2. In the specific employee profile table, data will be attached to an employee, who has been assessed (anassessee). All data will be stored at the lowest level, which means that obtained Likert value is related to the following information:

- Selected competence at level-3
- Date and time of the assessment
- Specific statements' text
- Specific value related to the statement
- The person who has given the assessment value

This process results in a table with data stored on raw data level. Therefore, it is possible to create competence analysis on several higher levels by specific calculations and statistical analysis. For each job profile, a set of relevant competences is related to the job. The set of competences is selected from the competence tree. Because the specific competence can be more or less important for the job function, a weighting mechanism, is adopted to specify a weight factor for each competence that belongs to the specific job profile. The HRM department in cooperation with the specific head of department maintains the weight factors, when the job is defined.

This process is suitable when an expression of the competence level of an employee or a group of employees (a department) is wanted (human capital). The specific job related competences are weighted by means of the matching weight factor. Higher weighted competence value shows a higher competence level for an employee.

## 3.2. Self-Assessment

A head of department defines statements for the level-3 sub-sub-competence in the PIS tree and initiates questionnaires for the self-assessment method. This assessment method is like an exam to test the competence level of an assessee. Each level-3 competence should have at least one specific questionnaire. A job profile which should be defined by HR department consists of required competence categories from level-3 of the PIS tree. Once a job profile is set, there will be a complete test material that candidates can participate in the test in order to measure their acquired competence level for specific job roles.

An assessee is responsible for providing a response to statements in the self-assessment method which reflects how knowledgeable an assessee for associated competences in the PIS tree is. This measures a person's intangible job knowledge, that cannot be assessed through certificates and/or documents. The statements in this method are multi-choice questions. Domain experts (mainly the head of the department) should prepare these statements for associated competences to the target job profile that an assessee is going to be evaluated for it.

The weighted arithmetic mean of final answers are calculated for one competence, and stored in the system for defining the competence level achieved through self-assessment of specific competence. The domain expert who is normally the head of department in organizations should also define the weight for any question in the test. The sum of weights must be equal to one. These weights are being used for identifying difficulties and also the importance of questions in assessing specific competence. Configuration and assignment of the self-assessment is only possible for level-3 competences. Higher level competences as stated earlier will be measured based on achieved final values for level-3 competences.

## 4. A Mathematical Model of Competence Assessment

The mathematical model of competence assessment can be described as a two-step process, (1) the Evaluation Phase (EP) and (2) the Assessment Phase (AP). In the EP, an Acquired Competence Data (ACD) of an assessee is being obtained which can be integrated into an employee profile in each HRM system. In some research works, the term "acquired competence" or "available competences" are used for ACD. Moreover, the ComProFITS system collects information related to the minimum competence requirements, the Requested Competence Data (RCD), which is necessary for assessing an employee in a position or for hiring a new staff from the applicants' repository. RCD is structured as a matrix which consists of the weights of competences required for a target job position. Its dimensions are same as the ACD matrix and is based on the reference competence pyramid. The RCD matrix is being initiated by the company. Central role in the competence assessment mechanism also has the Prioritization of Competences (PoC), a hierarchical process of assigning weights that define the importance and the relative mass of required competences for a specific job.

Regarding the mathematical description of the EP, a $n \times m_i$ matrix $\mathbf{A}^{(level_k)}$, contains the ACD scores $a_{ij}^{(level_k)}$ where $k = 1,2,3$ denotes the level of the competence hierarchy, $i = 1,...,n$ denotes the individuals being assessed (assessee, employee or applicant) and $j = 1,...,m_k$ denotes the competence category of level $k$. Each element of the ACD matrices is an assessment expressed by a mark between 1 and 5, i.e. $1 \leq a_{ii}^{(level_k)} \leq 5$. These marks can be transformed for the needs of calculations in later steps of analysis.

There are a total number of 48 competence definitions in level-3 ($m_3 = 48$), which are the initial data. These scores constitute the basis of the EP and can be used in turn to evaluate the scores of the upper level categories of the hierarchical organization of competences in the PIS tree (Figure 1). Similarly, $m_2 = 12$ for level-2 and $m_1 = 3$ for level-1 in the PIS tree. More precisely, the groups of items in the sub-sub categories (level-3) can be accumulated via a weighted schema providing the scores of level-2. Since each competence of level-2 contains four sub competences of level-3, then a generic formula of calculating competences of level-2 is:

$$a_{ij}^{(level_2)} = \sum_{r=1}^{4} w_{[4(j-1)+r]}^{(level3)} \cdot a_{i,[4(j-1)+r]}^{(level_3)} \tag{2}$$

$$0 \leq w_{[4(j-1)+r]}^{(level3)} \leq 1 \quad , \qquad \sum_{r=1}^{4} w_{[4(j-1)+r]}^{(level3)} = 1$$

where the weights, are predefined for every competence $j$ of the 2nd level. In the simplest case, all weights can be set equal to 1/4 and thus calculate the arithmetic mean of the 3rd level sub-competences. Similarly, with an analogous predefined weighing schema, the scores of level-2 provide the scores of level-1. For further details about measuring ACD for instance to the level-3, see Table 2. ACD for other levels in the PIS tree can be measured using the same method. Summarizing what we have described above, it is clear that the assessment procedure of competences is a bottom-up procedure that has to be initiated from the 3rd level of the hierarchy.

**Table 2.** Sample ACD formulas for level-3 in PIS competence tree ($m_3 = 48$)

| Person | $C_{1.1.1}$ | $C_{1.1.2}$ | … | $C_{3.4.4}$ |
|--------|-------------|-------------|-----|-------------|
| Person 1 | $a_{1.1}^{level_3}$ | $a_{1.2}^{level_3}$ | … | $a_{1.m_3}^{level_3}$ |
| Person 2 | $a_{2.1}^{level_3}$ | $a_{2.2}^{level_3}$ | … | $a_{2.m_3}^{level_3}$ |
| … | … | … | … | … |
| Person $n$ | $a_{n.1}^{level_3}$ | $a_{n.2}^{level_3}$ | … | $a_{n.m_3}^{level_3}$ |

The PoC is a weighting mechanism assigning weights to competences in order to express the relative preference of a specific competence with respect to others for a specific position. These weights can be assigned in every level and used in the later phases of decision making in combination with the competences of the other matrices, i.e. ACD and RCD. It is clear that the PoC is varied for different job positions. For the practical needs of ComProFITS, we assume that the weights are assigned only to level-1 and level-2.

For the PoC, we consider a voting scheme known as the Hierarchical Cumulative Voting (HCV) (Berander & Jonsson, 2006). The HCV is a version of the Cumulative Voting (CV) method, in which the assignment of the weights is based on the allocation of imaginary units in such a way that all amounts sum to a fixed number (say 100). In HCV the items to be assigned are structured into groups and the elements of the higher level (or groups) are assigned first amounts summing to 100 by CV and then the items on the lower level are prioritized separately within each group. In this manner, the PoC takes into account both the hierarchical organization of competences (Figure 1) and the quantitative prioritization in order to clearly demonstrate how much a specific competence is more important or relevant to a specific job than any other.

There are 3 tiers of items (Professional, Innovative and Social) in the PIS tree to be prioritized in the higher level (level-1). Let $C_i$ where $i = 1, 2, 3$ denotes the i-th tier (level-1) and that to each one of the groups the following amounts are assigned:

$$w_i, \quad 0 \le w_i \le f, \quad i = 1,2,3 \quad such\ that \quad \sum_{i=1}^{k=3} w_i = f \tag{3}$$

The sum of the allocated values is prefixed, it is usually set to f = 100 without loss of generality, since for any value of *f*, the weights can be easily transformed (by division by f) to have sum equal to 1. Since $C_i$ is a level-1 competence group containing 4 items of level-2, these have also to be prioritized in a similar way. So, for each group $C_i$, $i$ = 1, 2, 3, the prioritization of its 4 issues are denoted:

$$y_{ij}, \quad 0 \le y_{ij} \le f, \quad j = 1,2,3,4 \quad with \quad \sum_{j=1}^{4} y_{ij} = f \tag{4}$$

Schematically, the weights are assigned according to the hierarchical model shown in Figure 1. HCV is therefore a top-down approach that is initiated at level-1 and is terminated at level-2 of the hierarchy. As an example, for the level-1 categories, $C_1$ (Professional competences), $C_2$ (Innovative competences) and $C_3$ (Social competences), if the distribution of *f=100* is $w_1$ = 60, $w_2$ = 20 and $w_3$ = 20, then it is clear that $C_1$ is the most important competence category for the job and is considered three times more important than $C_2$ and $C_3$ competences which are considered equally important.

Continuing the example, in the second level, f=100 are distributed again at each sub-category of $C_{1.1}$, $C_{1.2}$, $C_{1.3}$ and $C_{1.4}$. If, for example, we distribute 100 in the four categories of C1 (e.g. managerial competences, business orientation, job related skills and oral and written communication/languages) as $y_{1,1}$ = 60, $y_{1,2}$ = 10, $y_{1,3}$ = 30 and $y_{1,4}$ = 20, then it is clear that managerial competences ($C_{1.1}$) are considered three times more important than written communication/languages ($C_{1.4}$) and six times more important than business orientation ($C_{1.2}$) and two times more important than job related skills ($C_{1.3}$).

It is important to note that if we combine ACD and PoC in the 2nd level, the values assigned to 2nd level should take into account the values assigned to 1st level. This can be achieved by a simple multiplication and a normalization which results in values summing to 1. The normalization and adjustment results in the absolute assignment of importance to each level-2 competence such that their sum regardless of category is 1.0. It is worth noting that the HCV schema has been used to requirements prioritization in information systems and in general produces vectors of very interesting and peculiar data (since they essentially are comprised of proportions summing to 1), which are suitable for advanced statistical methodologies (Compositional Data Analysis-CoDA) (Berander & Jonsson, 2006; Chatzipetrou, Angelis, Rovegard, & Wohlin, 2010).

The ComProFITS system combines the PoC data with both ACD and RCD scores in order to prioritize employees and new applicants with specialized competences (ACD) and give the proper significance to the new job under assessment (RCD). The combined data are the input to the following steps of the process which apply algorithmic approaches to determine the suitability of available resources (staff) with respect to the skills required for a specific job.

Except from the solution that allocates employees and applicants to specific job(s), the scores of ACD and RCD along with the PoC weights can be potentially used to derive meaningful conclusions and also visualize the results. Towards this direction, certain multivariate statistical approaches can be the basis for this specialized meta-analysis. For example, gap analysis tools and dashboards can be used for the evaluation of the actual competences of the staff and the potential competences needed for a specific company. They present an easy to read manner with certain performance indicators in order to enable instantaneous decisions that have to be made from the heads of the departments.

### 4.1. Gap function

Despite the fact that the weighted scores and the graphical inspection through plots are quite informative and provide a quick overview of the competences, they cannot be easily used for the assignment problem of the position. For this reason, we decided to utilize the notion of a loss or gap function (Nikolaos Mittas & Angelis, 2013a) in order to capture the deviance between the ACD and RCD scores.

Modifying the initial definition provided by Hernández-Orallo (Hernandez-Orallo, 2013), a gap function is a cost function $g : \mathbf{A} \times \mathbf{R} \rightarrow \Re$, which compares scores in the competence domain quantifying the gap between ACD and RCD scores. In our case, the first argument will be the actual competence score and the second argument will be the requested competence score, $g(a, r)$.

Inspired by regression-based problems, in which the objective is the characterization of performances of alternative prediction candidates (Mendoza, 2006), there are several gap functions that can be used for the assignment problem and typical examples are the simple gap (SG) with $g_i = A_i - R_i$, the absolute gap (AG) with $ag_i = \left| A_i - R_i \right|$ or even the squared gap (SQG) with $sqg_i = \left( A_i - R_i \right)^2$ (Hernandez-Orallo, 2013).

### 4.2. Qualification Space

Despite the need for well-established mathematical concepts that quantify the gap between the ACD and RCD scores, there is also an imperative need to develop and adapt highly readable and easily-interpretable visualization mechanisms, which will aid the head of the HR departments and managers to gain better insights of the assesses. Towards this direction, we propose the utilization of *Regression Operating Characteristic* (ROC) space (Hernandez-Orallo, 2013), which is initially proposed in order to graphically represent over-estimation against under-estimation of prediction models.

Based on the notion of RROC space, we can similarly define the *Qualification Space* (QS) as a two-dimensional plot, where the horizontal axis (or x-axis) represents the weighted sum of over-qualification (SOQ) and the vertical (or y-axis) represents the weighted sum of under-qualification (SUQ) for a specific candidate, calculated both using the simple gap function.

$$SOQ = \sum_{i=1}^{n} \left\{ g_i \mid g_i \geq 0 \right\} = \sum_{i=1}^{n} \left\{ w_i \left( A_i - R_i \right) \mid \left( A_i - R_i \right) \geq 0 \right\}$$

*(5)*

$$SUQ = \sum_{i=1}^{n} \left\{ g_i \mid g_i < 0 \right\} = \sum_{i=1}^{n} \left\{ w_i \left( A_i - R_i \right) \mid \left( A_i - R_i \right) < 0 \right\}$$

*(6)*

The relative position of a point offers additional information regarding the competences of the assessees, which can be over or under-qualified. In fact, all points above the diagonal line correspond to assessees with over-qualification, whereas the opposite is true for assessees lying below the diagonal. For example, the point visualizing the competences of assessee 2, indicates a person with an over-qualification. In contrast, another assessee can be considered with the least qualifications for the specific job, since the point that corresponds to its performance has the highest vertical distance from the reference line. Finally, another assessee seems to balance between over and under qualification. The visualization of this example is provided in Figure 2.

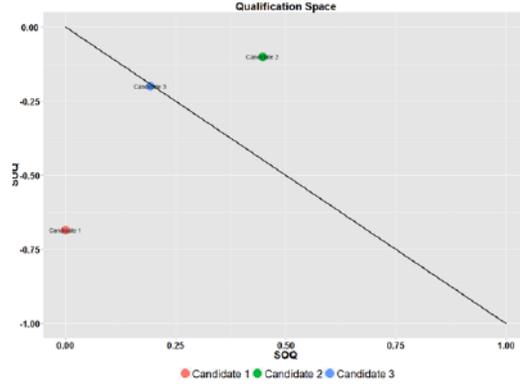

**Figure 2**. Qualification Space of the three over-, under-, and best-fit- candidates discussed as an example in the paper

Quantifying the visual information gained from the QS plot, we can easily define overall measures of gap for each assesse through the SOQ and SUQ values. For example, the Mean Simple Gap (MSG) indicator is the sum of SOQ and SUQ values divided by the number of competences of level-2. Similarly, the Mean Absolute Gap (MAG) is the sum of the absolute SOQ and SUQ values divided by the number of competences of level-2.

$$MSG = \frac{\sum_{i=1}^{n} g_i}{n} = \frac{SOQ + SUQ}{n} \qquad (7)$$

$$MAG = \frac{\sum_{i=1}^{n} ag_i}{n} = \frac{SOQ + |SUQ|}{n} \qquad (8)$$

An interesting geometrical property of QS is that the above mentioned indicators can be evaluated via a straightforward manner. More precisely, the sum of SG values (nominator of MSG equation) equals the length of the two parallel (horizontal and vertical) segments from the assessee point to the diagonal reference line (Figure 3a) (Hernandez-Orallo, 2013). Finally, the sum of AG values (nominator of MAG equation) is the Manhattan distance between the two-dimensional point and the origin point (0, 0) of QS (Figure 3b) (Hernandez-Orallo, 2013).

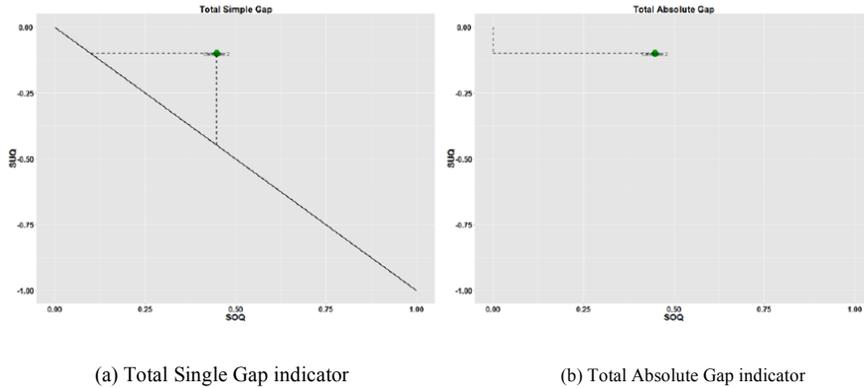

(a) Total Single Gap indicator                      (b) Total Absolute Gap indicator

**Figure 3.** Total Single and Absolute Gap Indicator plots for assessees of hypothetical scenario

## 4.3. Assessment Phase through a Multiple-Comparisons Algorithm

After the EP of the assesses, the objective of the AP is the ranking and clustering of the candidates according to their competence profile (ACD) and the competence requirements (RCD) of a specific position. Based on the notions described in the previous sections, we propose a formal framework, which is totally settled on well-established statistical principles, such as Design of Experiments (DOE) and multiple-hypothesis testing.

### 4.3.1. Design of Experiments

Describing briefly, DOE refers to a systematic planning in order to maintain control over all factors that may affect the result of an experiment, while it constitutes an entire branch area in statistics involving fundamental concepts that have to be specified and controlled in advance (Antony, 2003). The basic element of a DOE is the experimental unit which is the "object" on which the researcher wishes to measure a response variable. The purpose is to study the effect of one or more factors (categorical variables) on the response variable. The different categories of a factor are known as levels or treatments.

In our context, the competence profile of each candidate for a specific position is evaluated through the gap functions described in Section 4.1. More precisely, after the estimation of the weighted competences scores of ACD and RCD for level-2, we make use of the SG function, defined in section 4.2, to evaluate the gap between the weighted actual competences of candidates and the required competences of the job. The derived matrix describes essentially the DOE of the proposed methodology. Following the terminology of the DOE, the 12 different competences of level-2 can be considered as the experimental units in our context, the comparative candidates represent the treatments and the response variable is the weighted expression of the SG function, which represents the gap between the required and actual competences scores. The purpose is to investigate the effect of different treatments (candidates) on the response gap variable, i.e. to test the differences of the competences of comparative candidates.

Depending on the number of factors and the sources of variation that can be accounted for, on the response variable (SG), it is important to identify the most appropriate DOE technique that has to be adopted, i.e. the arrangement of the factors that will provide statistically valid results. Statistical literature offers a plethora of DOEs, which can be utilized in different problems under certain conditions. This is the reason why we decided to employ a specific type of DOE, namely the Repeated Measures Design (Mendoza, 2006) which is equivalent to the Randomized Complete Block Design (RCBD) (Higgins, 2003).

The RCBD setup incorporates an additional factor, the so-called block, which takes into account the grouping of similar experimental units. The incorporation of this extra factor is considered advantageous in order to identify true differences between treatments or equivalently the true treatment effect. When different treatments are applied to similar (or the same) experimental units (candidates in our context), which form in any sense a block, there is a source of variation between blocks which cannot be explained by the difference between treatments. This source of variation is represented by the block factor that is considered in the analysis. In this context, the sub-competences of level-2 represent the blocking factor. The adaptation of RCBD in our context can be described by the following equation

$$g_{ijk} = \mu + \alpha_t + \beta_j + \varepsilon_{ijk} \qquad (9)$$

where, $g$ s are the gap scores for each candidate, $\mu$ is the overall mean, $\alpha$ s are the treatment effects or the different competences of level-2 and $\beta$ s are the block effects or the candidates in our framework.

### 4.3.2. Ranking and Clustering Algorithm.

The central role to the proposed AP exhibits the utilization of a multiple comparison procedure, which is based on principles of cluster analysis and is used as the core methodology for the ranking and clustering of candidates, namely the Scott-Knott (SK) algorithm (Scott & Knott, 1974). The SK methodology presents a noteworthy property over other multiple hypothesis testing procedures, since it constructs homogeneous groups of candidates that are placed in non-overlapping groups, thus the output of the algorithm is not just one qualified candidate for a specific job but rather a cluster of candidates presenting statistically superior competence profiles than the rest. The proposed algorithm has been applied with success in the demanding task of the selection of the most accurate prediction model in Software Cost Estimation (SCE) research area (Nikolaos Mittas & Angelis, 2013b).

Describing briefly, the SK procedure utilizes the statistical method for Analysis of Variance (ANOVA) testing the null hypothesis that there is no statistical difference between the means of the gap scores obtained by the compared candidates. The contextualized null hypothesis is that the head of HR should not worry about the selection of the most appropriate candidate for a specific job, since all candidates present similar competence profiles. In contrast, the alternative hypothesis is that the candidates can be partitioned into two mutually exclusive and collectively exhaustive non-empty subsets. This practically means that there are candidates, who are more performant and competitive, according to their competences for the target job position than others and thus, the manager should take into account the extracted knowledge derived from the analysis of the competence gap.

The procedure we propose consists of consecutive steps aiming at a maximum differentiation between groups of candidates at each stage. Each group that is formed can be partitioned again if the new groups are significantly different. The whole methodology and associated algorithms are fully described in (Nikolaos Mittas & Angelis, 2013b) and presented in the following steps after the adaptation in the proposed framework:

1. Sort the means of the gap scores $\bar{g}_i, i = 1,...,n$ for each candidate in ascending order:

$$\bar{g}_{(1)} \leq \bar{g}_{(2)} \leq ... \leq \bar{g}_{(n)} \qquad (10)$$

2. For each $\bar{g}_{(i)}, i = 1,...,n-1$ separate the group of all ordered means $G$ into two subgroups $G_1 = \{\bar{g}_{(1)},...,\bar{g}_{(i)}\}$ and $G_2 = \{\bar{g}_{(i+1)},...,\bar{g}_{(n)}\}$ and compute the between-groups sum of squares:

$$SS_{BGi} = k\left( |G_1| (\bar{g}_{G_1} - \bar{g}_G)^2 + |G_2| (\bar{g}_{G_2} - \bar{g}_G)^2 \right) \qquad (11)$$

where $|G_1|, |G_2|$ are the cardinalities of the two subgroups and $\bar{g}_G, \bar{g}_{G_1}, \bar{g}_{G_2}$ are the means of groups $G, G_1$ and $G_2$, respectively:

$$\bar{g}_G = \frac{1}{d} \sum_{j=1}^{d} \bar{e}_{(j)},$$

$$\bar{g}_{G_1} = \frac{1}{|G_1|} \sum_{i \in G_1} \bar{g}_{(i)}, \quad \bar{g}_{G_2} = \frac{1}{|G_2|} \sum_{i \in E_2} \bar{g}_{(i)} \tag{12}$$

3. Find the partition that maximizes the value of the above sum of squares:

$$BG_{SS_i}* = \max\left\{BG_{SS_i}, \ i = 1, ..., n\right\} \tag{13}$$

4. Compute from the ANOVA table the $s^2$, the estimation of $\sigma^2$ (the variance that cannot be explained by the factors, i.e. the treatments and the blocks) by dividing the sum of squares by the corresponding degrees of freedom. Next, compute the statistic

$$\lambda = \frac{\pi}{2(\pi - 2)} \frac{BG_{SS_i}*}{s^2} \tag{14}$$

which has approximately a $\chi_v^2$ distribution where the degrees of freedom are given by $v = k/(\pi - 2)$ (rounded).

5. If $\lambda > \chi_{v;a}^2$ (where $\alpha$ is a predefined significance level), then the same test is applied to each group separately. If $\lambda < \chi_{v;a}^2$, then all means belong to the same homogeneous group. The procedure is continued by splitting each group into two subgroups if the $\lambda$-criterion is significant, otherwise by identifying a homogeneous group, until no groups can be split further.

Here, we have to point out that all the aforementioned mathematical and statistical techniques are applied to employees or candidates who are eligible according to certain predefined criteria. So, it is assumed that there is always a stage of eligibility check where candidates are excluded from the ranking process when their competences do not meet the minimum predefined criteria (for example a certain type of University degree).

## 5. Discussion on Practical and Test Results

In the context of the ComProFITS project, the final developed system is a Java Enterprise Edition web application. The application supports multiple roles: The head of department role, the administrator role, the generic employee role, various human resource management roles (recruiter, assistant and team development) and the job applicant role. Each role can perform several activities and some activities are provided in more than one role. In this section, the usage of the application for the purpose of the mathematical approach is concentrated to competence assessment described earlier in Section 4. Further practical and technical details of the ComProFITS software are described in (Mahdi Bohlouli, Ansari, Kakarontzas, & Angelis, 2015; N Mittas et al., 2015).

In order to assess an employee, the immediate manager can run a 360 degrees assessment process by forming a team of 5 assessors which also includes the assessee, by deciding which level-3 competences to assess and by selecting statements for assessing these competences (Figure 4) or the manager can assign competences directly (Figure 5).

**Figure 4.** 360 Degrees Assessment creation in the ComProFITS web application

**Figure 5.** Direct assignment of competence values for level-3 competences in the ComProFITS web application

In both cases level-3 competences will be given a specific value. In addition, job descriptions can also be stored in the system (Figure 6). In this process, the importance of a level-3 competence for a specific job position is specified. Possible values are "Very Important", "Important", "Moderately Important", "Of Little Importance" and "Unimportant". These lexical terms are mapped to a scale of 5, 4, 3, 2 and 1, respectively. For level-2 and level-1 competences, their relative weight in relation to this job position is given. The values provided for the competences in the same category and level must sum up to 100, as described earlier.

After employees and jobs have been assigned specific competence values for all level-3 competences, and after weights have been assigned to higher level competences of the jobs, the system provides the capability to apply the mathematical model of competence assessment described in Section 4 and produces a detailed PDF report for each employee. The PDF contains among other elements graphs comparing actual (ACD) vs. requested (RCD) competences for level-2 and level-3 competences, such as those depicted in Figure 1.

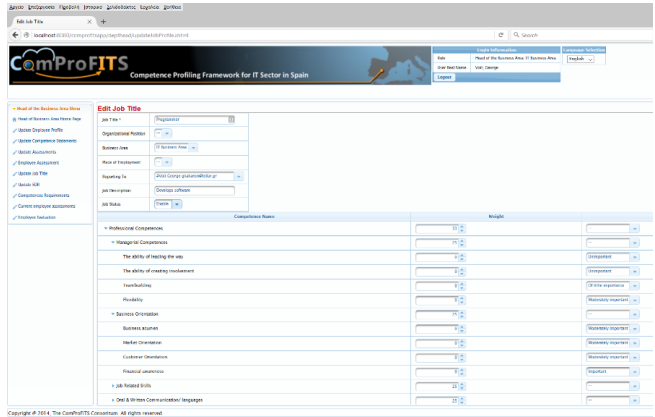

**Figure 6.** Providing requested competences for a job position in the ComProFITS web application

In the rest of this section, first we describe the data model in relation to the proposed mathematical model of assessment in Section 6.1 and the major software components involved in Section 6.2. Then, we discuss the empirical results of the validation of the ComProFITS application in an actual research center for the purpose of competence assessment of actual employees for a specific job position.

## 5.1. Data Model used for the implementation

A central notion crosscutting the ComProFITS project is the notion of competence assessment. Competences are organized in a hierarchical fashion as depicted in Figure 1. For example, the node numbered (1) in the tree in Figure 1, namely 'Professional Competences', is a level-1 competence and is calculated based on the values of its respective level-2 nodes, which are the nodes numbered 4, 5, 6 and 7. Each node at level-2 in turn, is calculated based on values from the respective sub-competences at level-3. For example, "Managerial Competences" are calculated based on values provided for the nodes numbered 16, 17, 18, and 19 at level-3. In order to implement this hierarchical competence structure in the data model, a single table related to itself suffices as depicted in Figure 7.

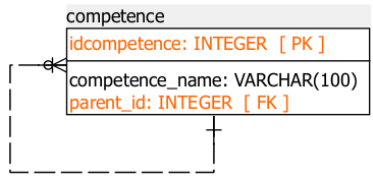

**Figure 7.** Competences data model

Competences table can be setup to contain the initial values depicted in Figure 1 as shown in Table 3.

**Table 3.** Initially provided values for the competences table (incomplete)

| idcompetence | Competence name | parent_id |
|---|---|---|
| 1 | Professional Competences | *NULL* |
| 2 | Innovative Competences | *NULL* |

| 3 | Social Competences | *NULL* |
| 4 | Managerial Competences | 1 |
| 5 | Business Orientation | 1 |
| 6 | Job Related Skills | 1 |
| 7 | Oral & Written Communication/ languages | 1 |
| 8 | Creativity and holistic thinking | 2 |
| 9 | Entrepreneurship | 2 |
| 10 | Proactivity | 2 |
| 11 | Readiness for changes | 2 |
| 12 | Teamwork | 3 |
| 13 | Professionalism | 3 |
| 14 | Interpersonal skills | 3 |
| 15 | Motivation for learning | 3 |
| 16 | The ability of leading the way | 4 |
| 17 | The ability of creating involvement | 4 |
| … | … | … |

We can then use this table to refer to the assessment of competences. For example, a value between 1 to 5, can be assigned to a competence assessment referring to entry 17 of the competences' table. This assessment regards '*The ability of creating involvement*' and will be used algorithmically for the assessment of its parent_id node which is '4' (i.e. Managerial Competences) along with other sibling values. Then, the value calculated for the node '4' can be used for the assessment of its parent_id node '1' (i.e. Professional competences) and so on.

In order to support the assessment of individual employees, two methods are supported. In the first, the employee (the employee entity in Figure 8) undergoes a 360 degrees' assessment process (the assessment entity in Figure 8) and the result of this assessment is recorded in the system (the employee_competence_assessment entity in Figure 8). In the second, the employee is assessed directly by his or her supervisor (the current_competence_assessment entity in Figure 8). All types of assessments, those derived from a 360 degree assessment by a team of five assessor employees or those assigned directly by a supervisor, refer to a specific competence (competence entity in Figure 8) and have a specific value which is the result of the assessment of the specific employee in the specific competence by a specific other employee who acts as an assessor (a member of the 360 degrees' assessment team or his/her supervisor). This result can be a value between 1 and 5. This is recorded in the assessment as an integer field in both cases of assessments (employee_competence_assessment and current_competence_assessment). Notice that the directly marked competences, this way are only level-3 competences in Figure 1. However, with the approach described in Section 4 higher level competences can be calculated algorithmically. These derived values are not recorded to avoid the danger of inconsistencies, since they can always be calculated from the low level competence assessments. The higher level values can also contain decimal digits (i.e. they are real numbers). In the simplest case, for example, consider all same level competences equally important, the higher level assessment based on these can be the average of these competence assessments.

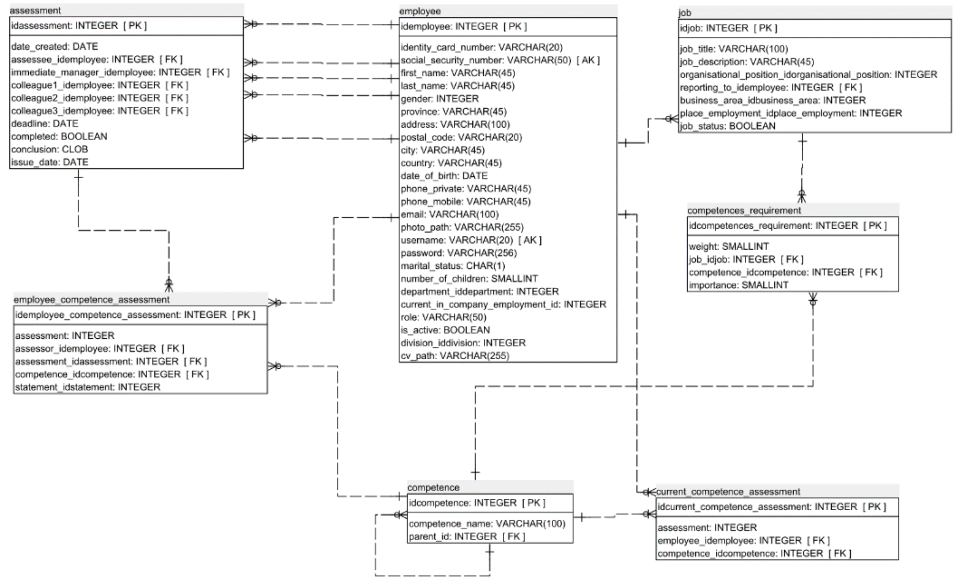

**Figure 8.** Part of the data model of ComProFITS application regarding employees' assessment

In order to calculate competence gaps (see Section 4) the required competences for a job are also recorded (competence_requirement entity in Figure 10). The competence requirements concern a specific job position (job entity in Figure 10) and a specific competence. Each job position carries a number of such competence requirements. For each competence requirement of a job, the importance and weight of this competence for this job position is recorded in the importance and weight fields of the competence_requirement entity in Figure 8, to enable the mathematical treatment described in Section 4.

The ComProFITS database comprises a total of 33 tables and 1 view and only a small part of this schema was described here for the purposes of better understanding the proposed mathematical approach and to illustrate better the support provided for the application of the proposed approach with the use of the ComProFITS web application the main elements of which are described in the next section.

## 5.2. ComProFITS software architecture

ComProFITS is a standard Java Enterprise Edition web application, which interfaces with the R language (R Developement Core Team, 2015) to perform the statistical method calculations. It follows closely the layered architectural style. The standard layered architecture approach requires a separation between the presentation layer, the business logic layer and the data handling layer. This provides several advantages including the possibility to distribute separate layers to distinct physical servers and therefore provide better scalability if necessary. The entities discussed earlier are mapped in java classes of the entities package which is accessed by the Java Server Faces (JSF) user interface via Enterprise Java Beans. The structure of the entities package is shown in Figure 9 below. The entities are separated into sub-packages according to the module that they belong. Figure 9 also depicts the dependencies among the various entities' packages.

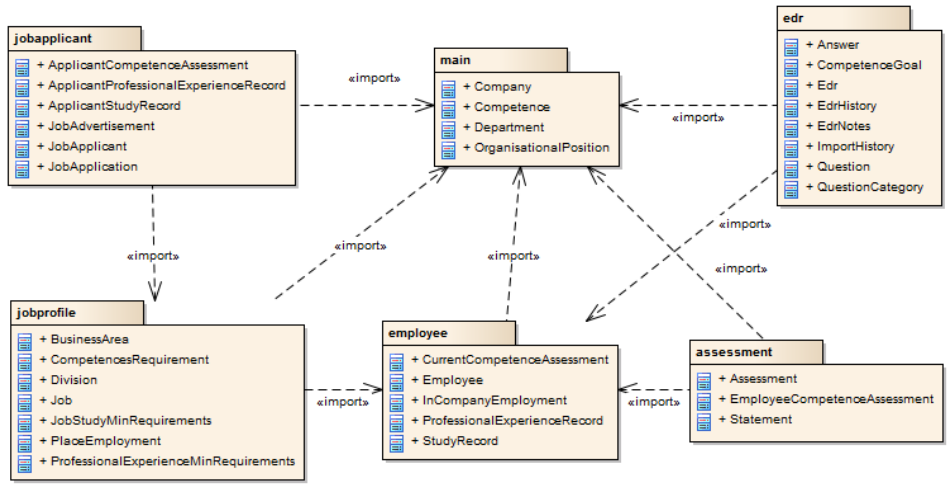

**Figure 9.** Packages of entities of the ComProFITS web application

In addition to the entities there are also a number of other packages and artifacts in the layered architecture. A high-level depiction of this architecture is presented in Figure 10. From Figure 10, the following layers have been developed:

1. ***UI Layer***: Where a number of pages and other artifacts serve the purpose of providing the user interface for the application. In ComProFITS, the Java Server Faces approach has been used for the user interface part, and more specifically the open source PrimeFaces component library.

2. ***Managed Beans***: A number of managed beans serving the purpose of the controller layer between the user interface and the rest of the application.

3. ***Session***: A number of enterprise java beans providing the application logic and the access to the database layer.

4. ***Entities***: The low level entities of the application such as employee and competence which in turn are directly mapped to tables in the database. For ComProFITS the Postrgresql database has been used. Although through the use of Java Persistence API any database specific code was avoided, so that the database should be relatively easy to change to a different technology.

### 5.3. Empirical validation of the mathematical approach to competence assessment

The results of this research were tested with real employee data shared from the project partners named Tecnalia. Tecnalia is a research center in northern Spain and as an end user of the project, has tested the system with over 200 employees' data and provided feedback for the scientific methods related to competence assessment and other aspects. In this way, the project outcomes directly contribute to the business processes at Tecnalia. Since the HRM data is connected with real employees, and should be analyzed confidentially, a Data Protection Agreement (DPA) was signed by project's end user. All employee data was anonymized to external project members to ensure compliance. Confidential data records, such as names were internally visible to Tecnalia members and externally anonymized for project members outside Tecnalia.

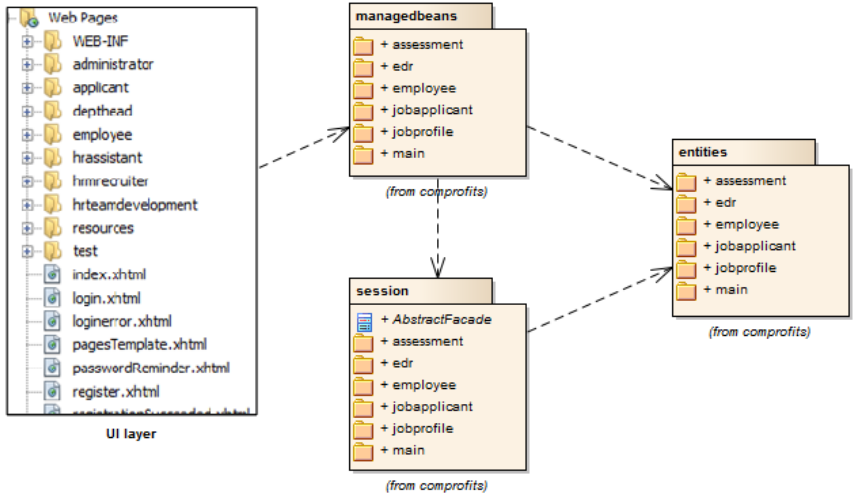

**Figure 10.** The layered architecture of the ComProFITS web application

For the goal of analyzing the practical results, specifically for the mathematical approach of competence assessment described in Section 4, we selected the real ACD data of 11 candidates as well as target job position definition as RCD. Further details of ACD and RCD data used in this case study are stated in Table 4. Furthermore, a head of department assigned the weights for levels-1 and level-2 of this specific job position profile using hierarchical cumulative Voting method (HCV) as shown in Figure 11.

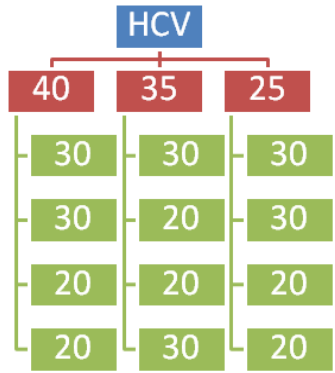

**Figure 11.** Hierarchical Cumulative Voting for competences of level-1 & level-2 in the case study of this research (TECNALIA)

Tables 5 and 6 represent the competency scores evaluated by the simple means of ACD derived from level-3, up to level-2 and level-1, respectively, whereas Figure 12 represent the results of ACD in level-1 for each candidate of the dataset. Furthermore, in Table 7, we present the overall descriptive statistical measures for the dataset and the distributions of competences for level-1 (Figure 13) and level-2 (Figure 14) of the competence tree.

**Table** 4. ACD and RCD of level-3 for eleven candidates (TECNALIA)

| Level 3 | Cnd 1 | Cnd 2 | Cnd 3 | Cnd 4 | Cnd 5 | Cnd 6 | Cnd 7 | Cnd 8 | Cnd 9 | Cnd 10 | Cnd 11 | Req |
|---|---|---|---|---|---|---|---|---|---|---|---|---|
| C1.1.1 | 4 | 1 | 3 | 4 | 4 | 1 | 3 | 4 | 4 | 5 | 4 | 5 |
| C1.1.2 | 4 | 1 | 3 | 4 | 3 | 1 | 2 | 3 | 4 | 4 | 4 | 5 |
| C1.1.3 | 4 | 1 | 2 | 5 | 4 | 1 | 3 | 3 | 4 | 5 | 3 | 3 |
| C1.1.4 | 3 | 1 | 4 | 4 | 4 | 2 | 3 | 3 | 4 | 3 | 4 | 2 |
| C1.2.1 | 4 | 2 | 3 | 4 | 3 | 1 | 2 | 3 | 3 | 4 | 4 | 4 |
| C1.2.2 | 4 | 1 | 3 | 4 | 3 | 1 | 3 | 3 | 3 | 4 | 4 | 4 |
| C1.2.3 | 4 | 2 | 4 | 5 | 4 | 1 | 3 | 3 | 3 | 4 | 4 | 3 |
| C1.2.4 | 3 | 1 | 3 | 3 | 2 | 1 | 3 | 2 | 3 | 3 | 3 | 3 |
| C1.3.1 | 3 | 2 | 3 | 4 | 4 | 1 | 3 | 3 | 4 | 3 | 5 | 4 |
| C1.3.2 | 4 | 3 | 3 | 3 | 4 | 2 | 2 | 4 | 3 | 4 | 3 | 4 |
| C1.3.3 | 3 | 1 | 3 | 3 | 4 | 1 | 3 | 3 | 3 | 3 | 3 | 3 |
| C1.3.4 | 4 | 3 | 3 | 4 | 3 | 2 | 2 | 3 | 3 | 3 | 3 | 4 |
| C1.4.1 | 4 | 1 | 3 | 4 | 4 | 2 | 2 | 3 | 4 | 3 | 3 | 3 |
| C1.4.2 | 3 | 1 | 2 | 5 | 2 | 1 | 4 | 3 | 4 | 4 | 4 | 4 |
| C1.4.3 | 4 | 1 | 3 | 5 | 2 | 2 | 4 | 3 | 4 | 4 | 3 | 4 |
| C1.4.4 | 4 | 1 | 2 | 4 | 4 | 1 | 3 | 4 | 3 | 4 | 3 | 3 |
| C2.1.1 | 3 | 3 | 3 | 4 | 4 | 2 | 3 | 4 | 4 | 4 | 4 | 4 |
| C2.1.2 | 4 | 3 | 3 | 5 | 4 | 1 | 3 | 3 | 4 | 3 | 3 | 3 |
| C2.1.3 | 3 | 2 | 3 | 4 | 3 | 1 | 3 | 3 | 3 | 3 | 4 | 3 |
| C2.1.4 | 4 | 2 | 3 | 4 | 3 | 1 | 3 | 3 | 3 | 3 | 3 | 4 |
| C2.2.1 | 4 | 3 | 3 | 5 | 4 | 1 | 2 | 3 | 4 | 4 | 4 | 4 |
| C2.2.2 | 3 | 2 | 4 | 5 | 3 | 1 | 2 | 3 | 3 | 3 | 3 | 4 |
| C2.2.3 | 3 | 2 | 3 | 4 | 4 | 1 | 2 | 4 | 3 | 3 | 3 | 4 |
| C2.2.4 | 3 | 2 | 3 | 4 | 3 | 1 | 3 | 3 | 3 | 3 | 4 | 3 |
| C2.3.1 | 3 | 2 | 2 | 4 | 3 | 1 | 3 | 2 | 3 | 4 | 4 | 4 |
| C2.3.2 | 4 | 2 | 3 | 5 | 4 | 1 | 2 | 3 | 3 | 3 | 4 | 4 |
| C2.3.3 | 4 | 1 | 3 | 4 | 4 | 1 | 3 | 4 | 4 | 5 | 3 | 4 |
| C2.3.4 | 4 | 1 | 3 | 3 | 4 | 1 | 2 | 4 | 4 | 5 | 4 | 4 |
| C2.4.1 | 4 | 2 | 3 | 4 | 4 | 2 | 2 | 4 | 4 | 3 | 3 | 4 |
| C2.4.2 | 4 | 2 | 3 | 4 | 4 | 2 | 4 | 3 | 4 | 4 | 3 | 3 |
| C2.4.3 | 3 | 3 | 3 | 3 | 4 | 1 | 2 | 4 | 3 | 3 | 3 | 4 |
| C2.4.4 | 3 | 1 | 3 | 3 | 3 | 1 | 3 | 3 | 4 | 4 | 3 | 3 |
| C3.1.1 | 5 | 1 | 2 | 4 | 4 | 1 | 3 | 4 | 4 | 5 | 4 | 4 |
| C3.1.2 | 5 | 1 | 3 | 4 | 5 | 1 | 3 | 5 | 4 | 5 | 3 | 3 |
| C3.1.3 | 5 | 1 | 2 | 4 | 5 | 1 | 3 | 4 | 4 | 5 | 3 | 4 |
| C3.1.4 | 4 | 1 | 2 | 4 | 3 | 1 | 3 | 4 | 4 | 4 | 3 | 3 |
| C3.2.1 | 3 | 2 | 3 | 4 | 3 | 1 | 3 | 4 | 3 | 4 | 4 | 3 |
| C3.2.2 | 3 | 1 | 4 | 4 | 4 | 1 | 2 | 4 | 4 | 4 | 5 | 4 |
| C3.2.3 | 4 | 3 | 3 | 4 | 4 | 1 | 3 | 4 | 3 | 4 | 3 | 4 |
| C3.2.4 | 4 | 1 | 3 | 5 | 4 | 1 | 2 | 4 | 4 | 5 | 3 | 3 |
| C3.3.1 | 4 | 1 | 3 | 5 | 5 | 1 | 3 | 4 | 4 | 5 | 3 | 4 |
| C3.3.2 | 3 | 1 | 3 | 4 | 4 | 1 | 3 | 3 | 4 | 4 | 2 | 4 |
| C3.3.3 | 3 | 1 | 4 | 4 | 5 | 2 | 4 | 3 | 4 | 3 | 3 | 3 |
| C3.3.4 | 4 | 1 | 3 | 4 | 4 | 1 | 3 | 4 | 4 | 4 | 3 | 4 |
| C3.4.1 | 4 | 3 | 3 | 3 | 4 | 1 | 2 | 4 | 3 | 3 | 3 | 4 |
| C3.4.2 | 4 | 3 | 3 | 4 | 4 | 1 | 2 | 3 | 3 | 3 | 3 | 4 |
| C3.4.3 | 3 | 3 | 4 | 4 | 3 | 1 | 2 | 3 | 3 | 3 | 3 | 3 |
| C3.4.4 | 3 | 3 | 4 | 3 | 3 | 1 | 2 | 3 | 3 | 4 | 3 | 3 |

**Table 5.** ACD and RCD of level-2 for eleven candidates (TECNALIA)

| Level 2 | Cnd 1 | Cnd 2 | Cnd 3 | Cnd 4 | Cnd 5 | Cnd 6 | Cnd 7 | Cnd 8 | Cnd 9 | Cnd 10 | Cnd 11 | Req |
|---------|-------|-------|-------|-------|-------|-------|-------|-------|-------|--------|--------|------|
| C1.1 | 3.75 | 1.00 | 3.00 | 4.25 | 3.75 | 1.25 | 2.75 | 3.25 | 4.00 | 4.25 | 3.75 | 3.75 |
| C1.2 | 3.75 | 1.50 | 3.25 | 4.00 | 3.00 | 1.00 | 2.75 | 2.75 | 3.00 | 3.75 | 3.75 | 3.50 |
| C1.3 | 3.50 | 2.25 | 3.00 | 3.50 | 3.75 | 1.50 | 2.50 | 3.25 | 3.25 | 3.25 | 3.50 | 3.75 |
| C1.4 | 3.75 | 1.00 | 2.50 | 4.50 | 3.00 | 1.50 | 3.25 | 3.25 | 3.75 | 3.75 | 3.25 | 3.50 |
| C2.1 | 3.50 | 2.50 | 3.00 | 4.25 | 3.50 | 1.25 | 3.00 | 3.25 | 3.50 | 3.25 | 3.50 | 3.50 |
| C2.2 | 3.25 | 2.25 | 3.25 | 4.50 | 3.50 | 1.00 | 2.25 | 3.25 | 3.25 | 3.25 | 3.50 | 3.75 |
| C2.3 | 3.75 | 1.50 | 2.75 | 4.00 | 3.75 | 1.00 | 2.50 | 3.25 | 3.50 | 4.25 | 3.75 | 4.00 |
| C2.4 | 3.50 | 2.00 | 3.00 | 3.50 | 3.75 | 1.50 | 2.75 | 3.50 | 3.75 | 3.50 | 3.00 | 3.50 |
| C3.1 | 4.75 | 1.00 | 2.25 | 4.00 | 4.25 | 1.00 | 3.00 | 4.25 | 4.00 | 4.75 | 3.25 | 3.50 |
| C3.2 | 3.50 | 1.75 | 3.25 | 4.25 | 3.75 | 1.00 | 2.50 | 4.00 | 3.50 | 4.25 | 3.75 | 3.50 |
| C3.3 | 3.50 | 1.00 | 3.25 | 4.25 | 4.50 | 1.25 | 3.25 | 3.50 | 4.00 | 4.00 | 2.75 | 3.75 |
| C3.4 | 3.50 | 3.00 | 3.50 | 3.50 | 3.50 | 1.00 | 2.00 | 3.25 | 3.00 | 3.25 | 3.00 | 3.50 |

**Table 6.** Actual competences scores of level-1 (TECNALIA)

| Candidates | Total | Professional (C1) | Innovative (C2) | Social (C3) |
|------------|-------|-------------------|-----------------|-------------|
| Cnd 1 | 3.67 | 3.69 | 3.50 | 3.81 |
| Cnd 2 | 1.73 | 1.44 | 2.06 | 1.69 |
| Cnd 3 | 3.00 | 2.94 | 3.00 | 3.06 |
| Cnd 4 | 4.04 | 4.06 | 4.06 | 4.00 |
| Cnd 5 | 3.67 | 3.38 | 3.62 | 4.00 |
| Cnd 6 | 1.19 | 1.31 | 1.19 | 1.06 |
| Cnd 7 | 2.71 | 2.81 | 2.62 | 2.69 |
| Cnd 8 | 3.40 | 3.12 | 3.31 | 3.75 |
| Cnd 9 | 3.54 | 3.50 | 3.50 | 3.62 |
| Cnd 10 | 3.79 | 3.75 | 3.56 | 4.06 |
| Cnd 11 | 3.40 | 3.56 | 3.44 | 3.19 |

**Table 7.** Descriptive statistics for Actual competence scores of level-1 (TECNALIA)

| Competences | $N$ | $M$ | $SD$ | $Mdn$ | $min$ | $max$ |
|-------------|-----|-----|------|-------|-------|-------|
| Total | 11 | 3.10 | 0.90 | 3.40 | 1.19 | 4.04 |
| Professional (C1) | 11 | 3.05 | 0.90 | 3.38 | 1.31 | 4.06 |
| Innovative (C2) | 11 | 3.08 | 0.83 | 3.44 | 1.19 | 4.06 |
| Social (C3) | 11 | 3.18 | 1.00 | 3.62 | 1.06 | 4.06 |

After the estimation of the weighted competences scores of ACD and RCD for level-2 (Table 8), we make use of the SG function to evaluate the gap between the weighted actual competences of candidates and the required competences of the job (Table 9). In our experimental design the C=12 different competences of level-2 can be considered as the experimental units in our context or the blocking factor, the comparative candidates represent the treatments and the response variable is the weighted expression of the SG function, which represents the gap between the required and actual competences scores. The purpose is to investigate the effect of different treatments (candidates) on the response gap variable, i.e. to test the differences of the competences of comparative candidates, through the SK algorithm.

**Table 8.** Weighted (PoC) ACD and RCD of level-2 for eleven candidates (TECNALIA)

| Level 2 | Cnd 1 | Cnd 2 | Cnd 3 | Cnd 4 | Cnd 5 | Cnd 6 | Cnd 7 | Cnd 8 | Cnd 9 | Cnd 10 | Cnd 11 | Req |
|---|---|---|---|---|---|---|---|---|---|---|---|---|
| C1.1 | 0.45 | 0.12 | 0.36 | 0.51 | 0.45 | 0.15 | 0.33 | 0.39 | 0.48 | 0.51 | 0.45 | 0.45 |
| C1.2 | 0.45 | 0.18 | 0.39 | 0.48 | 0.36 | 0.12 | 0.33 | 0.33 | 0.36 | 0.45 | 0.45 | 0.42 |
| C1.3 | 0.28 | 0.18 | 0.24 | 0.28 | 0.30 | 0.12 | 0.20 | 0.26 | 0.26 | 0.26 | 0.28 | 0.30 |
| C1.4 | 0.30 | 0.08 | 0.20 | 0.36 | 0.24 | 0.12 | 0.26 | 0.26 | 0.30 | 0.30 | 0.26 | 0.28 |
| C2.1 | 0.37 | 0.26 | 0.32 | 0.45 | 0.37 | 0.13 | 0.32 | 0.34 | 0.37 | 0.34 | 0.37 | 0.37 |
| C2.2 | 0.23 | 0.16 | 0.23 | 0.32 | 0.25 | 0.07 | 0.16 | 0.23 | 0.23 | 0.23 | 0.25 | 0.26 |
| C2.3 | 0.26 | 0.11 | 0.19 | 0.28 | 0.26 | 0.07 | 0.18 | 0.23 | 0.25 | 0.30 | 0.26 | 0.28 |
| C2.4 | 0.37 | 0.21 | 0.32 | 0.37 | 0.39 | 0.16 | 0.29 | 0.37 | 0.39 | 0.37 | 0.32 | 0.37 |
| C3.1 | 0.36 | 0.08 | 0.17 | 0.30 | 0.32 | 0.08 | 0.23 | 0.32 | 0.30 | 0.36 | 0.24 | 0.26 |
| C3.2 | 0.26 | 0.13 | 0.24 | 0.32 | 0.28 | 0.08 | 0.19 | 0.30 | 0.26 | 0.32 | 0.28 | 0.26 |
| C3.3 | 0.18 | 0.05 | 0.16 | 0.21 | 0.23 | 0.06 | 0.16 | 0.18 | 0.20 | 0.20 | 0.14 | 0.19 |
| C3.4 | 0.18 | 0.15 | 0.18 | 0.18 | 0.18 | 0.05 | 0.10 | 0.16 | 0.16 | 0.15 | 0.15 | 0.18 |

**Table 9.** Gap scores of level-2 for eleven candidates (TECNALIA)

| Level 2 | Cnd 1 | Cnd 2 | Cnd 3 | Cnd 4 | Cnd 5 | Cnd 6 | Cnd 7 | Cnd 8 | Cnd 9 | Cnd 10 | Cnd 11 |
|---|---|---|---|---|---|---|---|---|---|---|---|
| C1.1 | 0.00 | -0.33 | -0.09 | 0.06 | 0.00 | -0.30 | -0.12 | -0.06 | 0.03 | 0.06 | 0.00 |
| C1.2 | 0.03 | -0.24 | -0.03 | 0.06 | -0.06 | -0.30 | -0.09 | -0.09 | -0.06 | 0.03 | 0.03 |
| C1.3 | -0.02 | -0.12 | -0.06 | -0.02 | 0.00 | -0.18 | -0.10 | -0.04 | -0.04 | -0.04 | -0.02 |
| C1.4 | 0.02 | -0.20 | -0.08 | 0.08 | -0.04 | -0.16 | -0.02 | -0.02 | 0.02 | 0.02 | -0.02 |
| C2.1 | 0.00 | -0.11 | -0.05 | 0.08 | 0.00 | -0.24 | -0.05 | -0.03 | 0.00 | -0.03 | 0.00 |
| C2.2 | -0.04 | -0.11 | -0.04 | 0.05 | -0.02 | -0.19 | -0.11 | -0.04 | -0.04 | -0.04 | -0.02 |
| C2.3 | -0.02 | -0.18 | -0.09 | 0.00 | -0.02 | -0.21 | -0.11 | -0.05 | -0.04 | 0.02 | -0.02 |
| C2.4 | 0.00 | -0.16 | -0.05 | 0.00 | 0.03 | -0.21 | -0.08 | 0.00 | 0.03 | 0.00 | -0.05 |
| C3.1 | 0.09 | -0.19 | -0.09 | 0.04 | 0.06 | -0.19 | -0.04 | 0.06 | 0.04 | 0.09 | -0.02 |
| C3.2 | 0.00 | -0.13 | -0.02 | 0.06 | 0.02 | -0.19 | -0.08 | 0.04 | 0.00 | 0.06 | 0.02 |
| C3.3 | -0.01 | -0.14 | -0.03 | 0.03 | 0.04 | -0.13 | -0.03 | -0.01 | 0.01 | 0.01 | -0.05 |
| C3.4 | 0.00 | -0.03 | 0.00 | 0.00 | 0.00 | -0.13 | -0.08 | -0.01 | -0.03 | -0.01 | -0.03 |

Table 10 presents the results of the ANOVA procedure on which the multiple hypothesis testing of the SK algorithm was based. The table shows the significance ($p$-value) for the treatment (candidates) and block effects (sub-categories of level-2), as well as the practical importance of the results in the corresponding population of results through the eta-squared statistic. The latter provides additional information, since $p$-values alone may not be informative enough and thus, we take into account the strength of the differences and practical significance of the derived results. Finally, in order to estimate the effect size, which is a measure of the importance of a result in a population of results, we make use of the eta-squared statistic with the benchmark values of 0.01, 0.06 and 0.14 for small, medium and large effect size respectively.

From Table 10, the blocking effect is statistically significant, $F(11, 110) = 2.437$, $p < 0.001$. The practical meaning of this finding is that the sub-categories of level-2 add a significant variability that cannot be explained by the candidates. The candidates seem to present systematic differences across the sub-categories of level-2. Furthermore, the effect size (eta-squared statistic) of the phenomenon is characterized large, since the value is higher than the threshold value of 0.14.

More importantly, the findings of Table 10 for the treatment effect (candidates) indicate statistically significant differences in the mean values of SG scores (MSG), $F(10, 110) = 43.144$, $p < 0.001$, across the candidates with generally high effect size of the population. This practically means that there are candidates that are generally better in terms of qualifications or competences than comparable ones. So, our next objective is to rank them and select a candidate from a subset that forms the 'best' repository of candidates for this specific job.

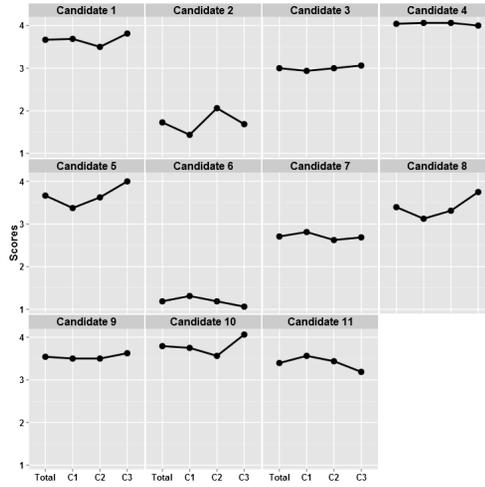

**Figure 12.** Actual competence scores of level-1 (TECNALIA)

Table 11 contains the output of the SK algorithm, in which the candidates are ranked according to their global qualifications starting from the best for the worst candidate (first column). The second column shows the mean values of the simple gap values (or MSG) for each candidate, whereas the third and fourth columns provide the MSG minus/plus one standard deviation. Finally, the derived clusters from the SK algorithm are colored with different shadings; candidates bearing the same coloring are clustered into the same homogenous group, indicating statistically similar mean gap values or in other words, similar qualifications.

Obviously, the sign of a MSG value indicates whether a candidate is over- or under-qualified for a specific job with positives (negative) values to reveal over- (under-) qualification. As far as the preference between an over-qualified candidate and a candidate with the smallest gap concerns, the selection is totally based on the policy of the organization/company.

In our example, the SK algorithm shows that the candidates can be grouped into five mutually exclusive clusters with similar qualifications. The findings can be easily summarized and graphically displayed through the Qualification Space (QS). Figure 15 represents the QS obtained from the gap values and the clustering algorithm executed on the dataset of the eleven candidates. The properties of QS (Section 4), all points above the diagonal line correspond to candidates with over-qualification, whereas the opposite is true for candidates lying below the diagonal. For example, the point visualizing the competences of candidate 4 indicates a person that is over-qualified. In contrast, candidate 6 can be considered as the candidate with the least qualifications for the specific job, from the point that corresponds to its performance has the highest vertical distance from the reference line. Finally, candidate 5 seems to find equilibrium between over and under qualification.

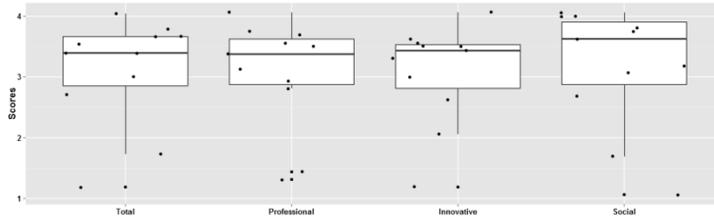

**Figure 13.** Distributions of competences of level-1

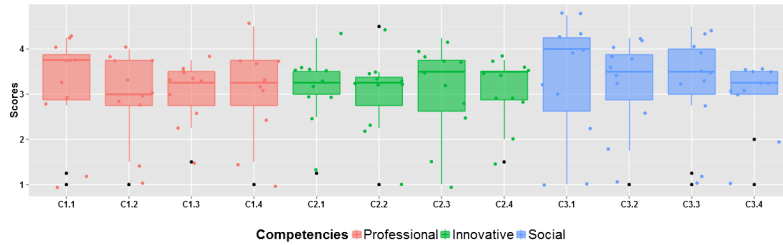

**Figure 14.** Distributions of competences of level-2

**Table** 10**.** Significance values and effect size (partial eta-squared) of ANOVA

| Factors | $df$ | $F$ | $p$ | Partial eta-squared |
|---|---|---|---|---|
| Candidates | 10 | 43.144 | < 0.001 | 0.797 |
| Sub-categories (level-2) | 11 | 2.437 | < 0.001 | 0.196 |

Concerning the findings of the clustering algorithm, the first cluster encompasses candidate 4 and candidate 10, which can be characterized as the most over-qualified candidates. They are candidates that present systematically higher competences than the requested for this specific job. The grouping into the same cluster practically means that there will be no difference, whether the head of the department will select either candidate 4 or candidate 10 for this specific job.

**Table** 11**.** Ranking and clustering of candidates (SK algorithm) (TECNALIA)

| Rank | candidates | MSG | Lower | Upper | Cluster | Qualification |
|---|---|---|---|---|---|---|
| 1 | candidate | 0.04 | 0.00 | 0.07 | candidate | Over- |
| 2 | candidate | 0.01 | -0.03 | 0.06 | candidate | Over- |
| 3 | candidate | 0.00 | -0.03 | 0.04 | candidate | Over- |
| 4 | candidate | 0.00 | -0.03 | 0.03 | candidate | Over- |
| 5 | candidate | -0.01 | -0.04 | 0.03 | candidate | Under- |
| 6 | candidate | -0.01 | -0.04 | 0.01 | candidate | Under- |
| 7 | candidate | -0.02 | -0.06 | 0.02 | candidate | Under- |
| 8 | candidate | -0.05 | -0.08 | -0.02 | candidate | Under- |
| 9 | candidate | -0.07 | -0.11 | -0.04 | candidate | Under- |
| 10 | candidate | -0.16 | -0.24 | -0.08 | candidate | Under- |
| 11 | candidate | -0.20 | -0.26 | -0.14 | candidate | Under- |

Most of the candidates (1, 5, 9, 11 and 8) are grouped into the second cluster, which is the cluster lying closest to the equilibrium line. So, if the policy of the company is to hire the most qualified candidates, then the head of the department should assign the job to one of the candidates (4 and 10) of the first cluster. On the other hand, if the policy of the company is to assign a specific job to the most appropriate candidate, according to the requested competences, then a manager should assign the job to a candidate of the second cluster representing the least gap between actual and requested competences. Finally, the third, fourth and fifth clusters encompass candidates that can be characterized as inappropriate for the forthcoming job. An idea for a future extension regarding the system would be to allow users to specify if the company prefers for a certain job position to hire someone overqualified if present or to hire someone closer to the equilibrium line. Then the ordering produced by the system could be provided in accordance with users' preferences. For example, instead of showing the most qualified employees first, employees closer to the requested competences could be presented first.

It is worth mentioning that the outcome of this experiment was evaluated by the manager who participated in the evaluation as accurate. In fact, the specific outcomes from the system (the generated PDFs for each individual employee) were provided to the manager and he was asked if he agreed or disagreed with the reported competence gaps. In all cases but one he agreed completely. The only case that he initially disagreed, was proven to be a data entry mistake and when corrected his agreement with the results of the algorithm was complete.

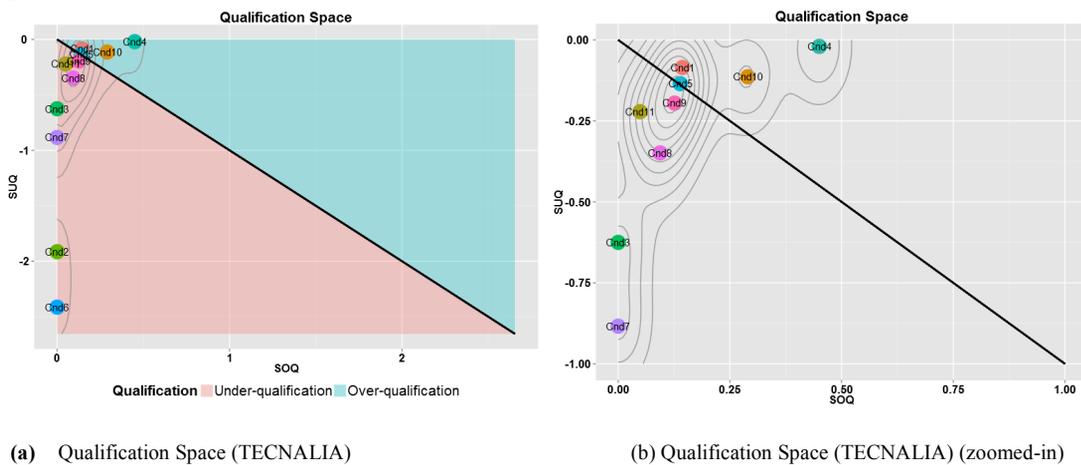

**(a)** Qualification Space (TECNALIA)          (b) Qualification Space (TECNALIA) (zoomed-in)

**Figure** 15**.** Qualification Space for the case study of this research from practical results

Concluding this section, we can say that HRM is a crucial activity for companies, especially for companies with a large number of employees. In such companies, it is not easy to maintain an accurate view of the employees' competences. The ComProFITS approach helps in maintaining this accurate view of employees' competences and based on job positions requested competences, it provides a mathematical approach for the assessment of competence gaps and for performing ranking and clustering. This helps companies to find the most suitable employees for a specific job position and employees to identify the most important gaps in their career development. In our experiments with Tecanalia, it was shown that the approach is very accurate in identifying the most suitable employees for a job position and for identifying competence gaps of existing employees. This is expected to have a largely positive impact in company's HRM approach. Therefore, there is a plan for the installation of ComProFITS in Tecnalia's premises and for using the application as a module integrated with their existing HRM system.

## 6. Conclusions and Outlook

Industries put significant effort to improve the quality of their products and services. This has a close connection with the job performance and an efficient use of human resources. Therefore, there is a need for proper competence analytics that uses scientific algorithms and methods efficiently in order to utilize human resources. The innovation that has been provided in this research covers statistical analysis of the competences in order to find the best fitting candidates for specific job positions in companies. The Scott-Knott clustering algorithm that has been applied in this work classifies job seekers into groups such as under-, over-qualified or best-fit candidates with respect to the specific job definition. Previously developed algorithms in this work have been evaluated and tested with real HR data. The industrial partner which has been involved in this research project has already validated and confirmed the accurate results achieved in the test phase.

From the sort of different assessment types as well as various competence categories in the reference competence tree, one may think that doing all these assessments for all of those stated competences in the tree is a time-consuming task and may not worth the benefit obtained from it. But the fact is that, the assessments of one specific person are to be handled once at the beginning and achieved results are used many times in different recruitments and job applications. In addition, one specific job area does not need to assess all competence categories stated in the tree. Furthermore, some of those competence categories such as university degrees or spoken languages can be proven from certificates without any need to assess them again.

As a next step and future work, there is a need to apply developed algorithms in other contexts and case studies, such as medical science and for assessment of medical experts in order to test it in further domains, to ensure the easy adaptation of the platform in a wide variety of sectors (Mahdi Bohlouli, Uhr, Merges, Hassani, & Fathi, 2010). In addition, the concept of collective competence analysis based on the competence measures of individuals is also very interesting topic as an extension of this work for further research. The application of social media streaming and also utilization of big data technology (M. Bohlouli, Dalter, Dornhofer, Zenkert, & Fathi, 2015) is a must in this regard. As a great extension of this work, a cloud deployed ComProFITS system facilitates the accessibility and availability of such system in the frame of XaaS (Keshavarzi, Haghighat, & Bohlouli, 2013).

## Acknowledgements.

Parts of the research presented in this paper has been funded with support from the European Commission under the grant no. DE/13/LLP-LdV/TOI/147642. This publication reflects the views only of the authors, and the Commission cannot be held responsible for any use which may be made of the information contained therein. The authors would like to thank all the participants of the ComProFits project for their positive contribution to the successful completion of the project. In addition, authors would like to thank Dr. Scott Harrison for his proofreading supports and constructive comments to improve the quality of this publication.

Mahdi Bohlouli works as a computer scientist at the University of Siegen, Germany. He is responsible for the large scale data analysis (Big Data), cloud computing and competence management areas at the institute of Knowledge Based Systems (KBS). Mahdi is a contact person and a member of the project management board for European projects called ComProFITS (www.comprofits.eu) and COMALAT (http://www.comalat.eu). At the same time, he served a program committee (PC) member of highly qualified conferences such as ISC2015, ADBIS15, EINS15 as well a peer reviewer for journals like Pattern Recognition Letters (Elsevier), Information Processing Letters (Elsevier) and Organizacija (Organization - Journal of Management, Information Systems and Human Resources.

Nikolaos Mittas received the BSc degree in mathematics from the University of Crete and the MSc and PhD degrees in informatics from the Aristotle University of Thessaloniki (A.U.Th). The main part of his publications in journals and conference proceedings is focused on the development and application of statistical, machine learning and data mining algorithms for analyzing data from the fields of Information Systems and Software Engineering. He works as adjunct faculty at the Computer Science Department of A.U.Th and Open University of Cyprus. Furthermore, he has also participated in several funded R&D Greek or European projects as a researcher and free-lancer of statnous (www.statnous.com) for academic and research institutes.

George Kakarontzas is an Assistant Professor at the Department of Computer Science and Engineering at T.E.I. of Thessaly. He holds a university degree in Informatics, an MSc in Object-Oriented Software Technology and a PhD in Quality Assurance of Component-Based Software Systems. George's research interests include Software Architecture, Component-based Software Engineering, Software Quality and Distributed Computing and publishes regularly in international conferences and journals. George participated in several projects funded both by national as well as EU resources. He serves as a PC member in several conferences and workshops and reviewer for articles for journals in the area of software engineering..

Lefteris Angelis studied Mathematics and received his Ph.D. degree in Statistics from Aristotle University of Thessaloniki (A.U.Th.). He is currently an Associate Professor at the Department of Informatics of A.U.Th and coordinator of the STAINS (Statistics and Information Systems) research group. His research interests involve statistical methods with applications in information systems and software engineering, computational methods in mathematics and statistics, planning of experiments and simulation techniques.

Theodosios Theodosiou received a BSc degree in Biology from Aristotle University of Thessaloniki. He also received an MSc degree in Informatics from University of Edinburgh with a specialty in Bioinformatics. His doctoral dissertation has the title "Statistical multivariate methods for data mining data from biological documents and ontologies" covering the wider area of biomedical text mining. The main part of his publications in journals and conference proceedings is focused on the development and application of statistical methods and models for analyzing large sets of biomedical data from documents, genes and proteins. Finally, he has participated in several funded R&D Greek or European projects (indicatively, BioSapiens – A European network for integrated genome annotation, CADSES, etc.) and is co-founder of statnous (www.statnous.com), an analytics startup company.

Madjid Fathi is a professor and vice-chair of the EECS Department at the University of Siegen, Germany. His research interests are focused on Knowledge Based System (KBS), knowledge management applications in medicine and engineering, knowledge transfer, organizational learning, and Knowledge Discovery from Text (KDT). He is the editor of "Integrated Systems: Innovations and Applications (2015)", "Integration of Practice-Oriented Knowledge Technology" (2013) and "Integrated Systems, Design and Technology" (2011) published by Springer. He, with his students, has published with more than 200 publications, including 25 Journal publications, and obtained four paper awards. He is a senior member of the IEEE as well as a member of the editorial board of five respective journals.